\documentclass{emulateapj}
\usepackage[figuresright]{rotating}

\shorttitle{Properties of M31 globular clusters}
\shortauthors{Wang et al.}

\begin{document}

\slugcomment{AJ, in press}
\title{STRUCTURAL PARAMETERS FOR GLOBULAR CLUSTERS IN THE OUTER HALO OF M31}
\author{Song Wang,\altaffilmark{1,2,3}  Jun
  Ma\altaffilmark{1,3} }

\altaffiltext{1}{National Astronomical Observatories, Chinese Academy
  of Sciences, Beijing, 100012, China;\\ majun@nao.cas.cn}

\altaffiltext{2}{Graduate University, Chinese Academy of Sciences,
  Beijing, 100039, China}

\altaffiltext{3}{Key Laboratory of Optical Astronomy, National Astronomical Observatories,
 Chinese Academy of Sciences, Beijing, 100012, China}

\begin{abstract}
In this paper, we present internal surface brightness profiles, using images in the F606W and F814W filter bands observed with the Advanced Camera for Surveys on the {\it Hubble Space Telescope}, for ten globular clusters (GCs) in the outer halo of M31. Standard King models are fitted to the profiles to derive their structural and dynamical parameters. The results show that, in general, the properties of clusters in M31 and the Milky Way fall in the same regions of parameter spaces. The outer halo GCs of M31 have larger ellipticities than most of GCs in M31 and the Milky Way. Their large ellipticities may be due to galaxy tides coming from satellite dwarf galaxies of M31 or may be related to the apparently more vigorous accretion or merger history that M31 has experienced. The tight correlation of cluster binding energy $E_b$ with mass $M_{\rm mod}$ indicates that, the ``fundamental plane'' does exist for clusters, regardless of their host environments, which is consistent with previous studies.

\end{abstract}


\section{INTRODUCTION}
\label{Introduction.sec}

The mechanisms involved in galaxy formation is still one of the major unsolved problems in astrophysics \citep[e.g.,][]{per02}. Globular clusters (GCs), which are considered to be debris of the galaxy formation, have a record about the information on both their formation condition and dynamical evolution within the environment of their host galaxies, which are reflected by their spatial structures and kinematics \citep{barmby07,maclau08}. So, GCs are regarded as a laboratory of galaxy history \citep{bs06}. In addition, GCs can be used as one of excellent tracers of substructures in the outer regions of their parent galaxies. For example, \citet{bfi} identified the accretion signature of the Sagittarius dwarf galaxy among the GCs in the outer halo of the Milky Way (MW); \citet{mackey07} found some of the GCs in the outer halo of M31 are rather unlike their MW counterparts as they are metal-poor, compact, and very luminous, which may well offer important clues to differences in the early formation and evolution of the two galaxies or in their subsequent accretion histories \citep[see][for details]{mackey07}. Thus, a detailed study of GCs in the outer halo of a galaxy is important.

M31, with a distance of $\sim780$ kpc from us \citep{sg98,mac01}, is the largest galaxy in the Local
Group, and it is so close to us that the GCs in it can be well resolved with the cameras on the {\it Hubble Space Telescope} ({\it HST}). M31 contains more GCs than all other Local Group galaxies combined with 654 confirmed GCs and 606 GC candidates in the version V4.0 of the Revised Bologna Catalogue (RBC) of M31 GCs \citep{gall04,gall06,gall07,gall09}.
M31 contains so many GCs that a variety of clusters may be included such as classic globulars, extended and diffuse globulars \citep{huxor08}, intermediate-age globulars
\citep{puzia05,fan06,ma09,wang10} and young massive clusters \citep{perina09,perina10,maetal11}.
GCs in the outer halo of M31 have been discussed by many authors
\citep{martin06,mackey06,mackey07,mackey10,huxor08}, which may provide important clues for the accretion and interaction events between M31 and surrounding galaxies.
Recently, \citet{mackey10} found a genuine physical association between GCs and multiple tidal debris streams in the outer regions of M31, implying that the remote GC system of M31 was largely accreted from the satellite galaxies \citep[also reported in][]{huxor11}.

Structures and kinematics of GCs can be determined by fitting different models to the surface brightness profiles, combined with mass-to-light ratios ($M/L$ values) estimated from velocity dispersions or
population-synthesis models. In general, three models are used in the fits: the simple model of
single-mass, isotropic, modified isothermal sphere developed by \citet{king66}, an alternate
modified isothermal sphere based on the ad hoc stellar distribution function of \citet{wilson75}, and
the $R^{1/n}$ surface-density profile of \cite{sersic68}. With these models, many authors have achieved
some success in determining structures and kinematics of clusters from different galaxies, using images from ground-based telescopes or {\it HST}: the MW \citep{tdk93,tkd95,mm05}; the Large and Small Magellanic Clouds, Fornax and Sagittarius dwarf spheroidal galaxies \citep{mg03a,mg03b,mg03c,mm05}; M31 \citep{grill96,bhh02,barmby07,barmby09,ma11,ma06,ma07,ma12,federici07,str09,huxor11}; M33 \citep{larsen02}; NGC 5128 \citep{hch99,harris02,mh04,maclau08}. A number of studies focusing on the correlations between cluster parameters have been performed, which showed that there exists a fundamental plane among most clusters, regardless of their different ``growing environment'' in different host galaxies.

\citet{barmby07} derived structural parameters for 34 GCs in M31 based on {\it HST} Advanced Camera for Surveys (ACS) observations, and the derived structural parameters were combined with corrected versions of those measured in an earlier survey in order to construct a comprehensive catalog of structural and dynamical parameters for 93 M31 GCs. \citet{barmby09} measured structural parameters for 23 bright young clusters in M31 based on the {\it HST} Wide Field Planetary Camera 2 (WFPC2) observations, and suggested that on average they are larger and more concentrated than typical old clusters. However, the sample clusters from \citet{barmby07} and \citet{barmby09} lie at projected radii $R_{p}<20$ kpc except for five clusters G001, G002, G339, G353 and B468, the projected radii of which are 34.55, 33.62, 28.68, 26.32 and 20.05 kpc, respectively; and most of the sample clusters lie at projected radii $R_{p}<10$ kpc. So, structural parameters for GCs in the outer halo of M31 are worthwhile to be determined. In addition, \citet{huxor11} derived structural parameters for 13 extended clusters (ECs) in the halo regions of M31 by fitting the \citet{king62} profiles to the photometry data taken with the Wide Field Camera on the Isaac Newton Telescope (INT) and MegaCam on the Canada-France-Hawaii Telescope, which may provide an interesting comparison with the structural parameters for GCs in the outer halo of M31.

In this paper, we determined spatial structures and kinematics for ten GCs in the outskirts of M31.
In Section 2, we give observations of the sample GCs and the data-processing steps to derive their surface brightness profiles. In Section 3, we determine structures and kinematics of the sample clusters with the model fitting. In Section 4, we discuss correlations of the structural and kinematic parameters of the sample clusters here combining with those of the Galactic and M31 clusters studied by other authors. Finally, we give our summaries in Section 5.

\section{DATA AND ANALYSIS METHOD}
\label{data.sec}

\subsection{Globular Cluster Sample}

As mentioned in the introduction, a detailed study of GCs in the outer halo of a galaxy is important, since they can serve as one of excellent tracers of substructures in the outer regions of their parent galaxy. Till now, for M31, most of the clusters whose structural parameters have been determined, lie at projected radii $R_{p}<10$ kpc. So, structural parameters for GCs in the outer halo of M31 are worthwhile to be measured. In this paper, detailed studies of the structures of a sample of ten GCs in the outer halo of M31 from \citet{mackey07} will be presented. These sample halo GCs are interesting. For example, \citet{mackey07} found some of them are rather unlike their MW counterparts as they are metal-poor, compact, and very luminous \citep[see][for details]{mackey07}. Eight of the ten halo GCs lie at projected radii $R_{p}>30$ kpc, of which two lie at very large distances from M31: $R_{p}\sim 78$ and 100 kpc, respectively. In addition, \citet{mackey07} estimated their metallicities, distance moduli and reddening values by fitting the Galactic GC fiducials from \citet{brown05} to their observed color-magnitude diagrams in the F606W and F814W filters of deep images observed with the the ACS Wide Field Camera (WFC) under the {\it HST} program GO-10394 (PI: Tanvir). This program was aimed to obtain deep high resolution photometry of outer halo GCs in M31 to study their stellar populations, line-of-sight distances and structural parameters. Targets were imaged in the F606W and F814W filters for $\sim$~1800s and $\sim$~3000s, respectively, with small dithers between various sub-exposures \citep{rich09}.

\subsection{Surface Brightness Profiles}
We used the analogous procedure adopted by \citet{barmby07} to produce surface brightness profiles with {\sc ellipse} in IRAF. The center positions of these clusters were determined by centroiding. Elliptical isophotes were fitted to the observed data, with no sigma clipping. Two passes of {\sc ellipse} task were run in the procedure. In the first pass, ellipticity and position angle (P.A.) were allowed to vary with the isophote semimajor axes; in the second pass, surface brightness profiles were derived on fixed, zero-ellipticity isophotes, meaning that we always had circularly symmetric intensity profiles, which would be fitted with circular structure models. The overall ellipticity and position angle were determined by averaging the {\sc ellipse} output over the isophotal semimajor axes, and the uncertainty is $\sigma$ .
Table 1 lists the average ellipticity, P.A. and some additional integrated data for the sample GCs. $VI$ magnitudes of 9 GCs and $I$ magnitude of GC6 are from \citet{huxor08}, while $V$ magnitude of GC6 is from \citet{rhh92}. The galactocentric distances, distance moduli, reddening values and metallicities are from \citet{mackey07}, while the uncertainties of [Fe/H] are assumed to be 0.6 as \citet{bh00} suggested for the standard deviation of the metallicity distribution of M31 GC system.

Raw output from package {\sc ellipse} is in terms of counts s$^{-1}$ pixel$^{-1}$, which needs to multiply by 400 to convert to counts s$^{-1}$ arcsec$^{-1}$, since the ACS/WFC spatial resolution is 0.05 arcsec pixel$^{-1}$. For drizzled ACS data, the units of counts are ELECTRONS (ACS Handbook). Two formulas were used to transform the ACS counts to surface brightness calibrated on the {\sc vegamag} system (ACS Handbook),

$\mu_{\rm F606W}/{\rm mag~arcsec^{-2}=26.398-2.5 \log(counts~s^{-1}}$
\begin{equation}
{\rm arcsec^{-1})},
\end{equation}
$\mu_{\rm F814W}/{\rm mag~arcsec^{-2}=25.501-2.5 \log(counts~s^{-1}}$
\begin{equation}
{\rm arcsec^{-1})}.
\end{equation}

However, occasional oversubtraction of background during the multidrizzling in the automatic reduction pipeline leads to ``negative'' counts in some pixels, so we worked in terms of linear intensity instead of surface brightness in magnitudes. Given ${M_{\odot,{\rm F606W}}} = +4.64$,
${M_{\odot,{\rm F814W}}} = +4.14$\footnote{See http://www.ucolick.org/$\sim$cnaw/sun.html.},
equations for transforming counts to surface brightness
in intensity were derived \citep[also see][for details]{barmby07},
\begin{equation}
I_{\rm F606W}/L_{\odot}~{\rm pc^{-2}\simeq0.8427\times(counts~s^{-1}~arcsec^{-1})},
\end{equation}
\begin{equation}
I_{\rm F814W}/L_{\odot}~{\rm pc^{-2}\simeq1.2147\times(counts~s^{-1}~arcsec^{-1})}.
\end{equation}

Table 2 gives the final, calibrated intensity profiles for the ten clusters but with no extinction corrected. The reported F606W- and F814W-band intensities are calibrated on the {\sc vegamag} scale. Column (7) gives a flag for each point, which has the same meaning as \citet{barmby07} and \citet{maclau08} defined.

\subsection{Point-spread Function}

As noted by \citet{barmby07} and \citet{maclau08} that, though the sample GCs here are well resolved with ACS/WFC, the core structures are still influenced by the point-spread function (PSF). We convolved the structural models developed by \citet{king66} (hereafter `King model') with a simple analytic description of the PSF before doing the model fitting, as given in \citet{barmby07},
\begin{equation}
I_{\rm PSF, F606W}/I_{\rm 0} = [1 + (R/0.0686~{\rm arcsec})^3]^{-3.69/3.0},
\label{eq:psf1}
\end{equation}
and
\begin{equation}
I_{\rm PSF, F814W}/I_{\rm 0} = [1 + (R/0.0783~{\rm arcsec})^3]^{-3.56/3.0},
\label{eq:psf2}
\end{equation}
with FWHMs of $0.125~{\rm arcsec}$ and $0.145~{\rm arcsec}$ in the F606W and F814W filters, respectively.

\subsection{Extinction and Magnitude Transformation}

When we fit models to the brightness profiles of the sample clusters, we will correct the inferred intensity profiles for extinction. The effective wavelengths of the ACS F606W and F814W filters are $\lambda_{\rm eff}\simeq$ 5918 and 8060 {\AA} \citep{siri05}. With the extinction curve $A_{\lambda}$ taken from \citet{car89} with
$R_V=3.1$, two formulas for computing ${A_{\rm F606W}}$ and ${A_{\rm F814W}}$ are derived: ${A_{\rm F606W}}\simeq2.8~E_{B-V}$; ${A_{\rm F814W}}\simeq1.8~E_{B-V}$. In addition, for easy comparison with catalogs of the GCs in the MW (see Section 4 for details), we transform the ACS/WFC magnitudes in the F606W filter to the standard $V$. \citet{siri05} has given transformations from WFC to standard $BVRI$ magnitudes both on observed and synthetic methods (see their Table 22). As synthetic transformations are based on larger color range and more safely employed, they should be considered the norm, unless some indicated cases \citep{siri05}. We used the synthetic transformation from F606W to $V$ magnitude
both on the {\sc vegamag} scale with a quadratic dependence on dereddened $(V-I)_0$. With the magnitudes in $V$ and $I$ bands and reddening values listed in Table 1, we found the $(V-I)_0$ values of all the sample clusters are larger than 0.4. So, the following transformation formula was applied here,

$(V-{\rm F606W})_0=-0.067+0.340(V-I)_0-0.038$
\begin{equation}
(V-I)_0^2,
\end{equation}
for which we estimated a precision of about $\pm 0.05$ mag.

\section{MODEL FITTING}
\label{model.sec}

\begin{figure*}
\figurenum{1}
\centering
\includegraphics[angle=0,scale=0.7]{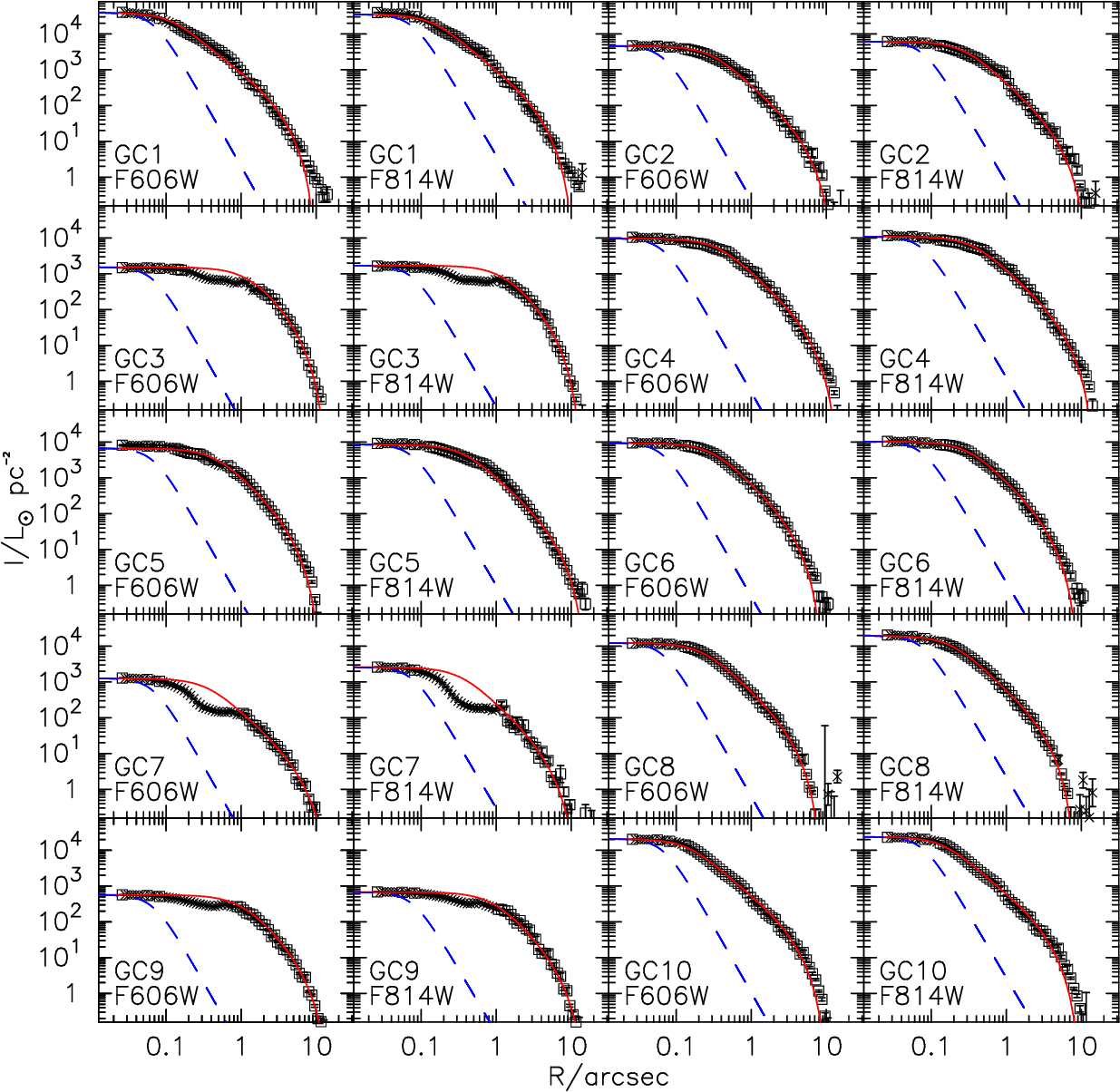}
\begin{minipage}[]{145mm}
\caption{Surface brightness profiles and model fits for the sample GCs here.}
\end{minipage}
\label{fig1}
\end{figure*}

There are a number of possible choices of structural models for fitting star cluster surface profiles, including King model, \citet{wilson75}, and \cite{sersic68}, as mentioned in the introduction. King model is the most commonly used model in studies of star clusters. In addition, \citet{barmby07, barmby09} found that M31 clusters are better fitted by King models. So, in this paper, the intensity profiles of the ten GCs in M31 will be fitted by King models defined by the phase-space distribution function,
\begin{equation}
f(E) \propto\left\{
\begin{array}{lcl}
\exp[-E/{\sigma}_0^2]-1, &      & E < 0, \\
0,                       &      & E \geq 0,
\end{array}\right.
\end{equation}
where $E$ is the stellar energy, ${\sigma}_0$ is a velocity scale.

We first convolved King model with the ACS/WFC PSF for the F606W and F814W filters. Given a value for the scale radius $r_0$, we computed a dimensionless model profile $\widetilde{I}_{\rm mod}\equiv I_{\rm mod}/I_0$, and then carried out the convolution,

$\widetilde{I}_{\rm mod}^{*} (R | r_0) = \int\!\!\!\int_{-\infty}^{\infty}
               \widetilde{I}_{\rm mod}(R^\prime/r_0)
               \widetilde{I}_{\rm PSF}
               \left[(x-x^\prime),(y-y^\prime)\right]$
\begin{equation}
       \ dx^\prime \, dy^\prime\ ,
\label{eq:convol}
\end{equation}
where $R^2=x^2+y^2$, and $R^{\prime2}=x^{\prime2}+y^{\prime2}$; and $\widetilde{I}_{\rm PSF}$ was approximated using
the equations (\ref{eq:psf1}) and (\ref{eq:psf2}) \citep[see][for details]{maclau08}. The observed surface brightness profiles were fitted by calculating and minimizing $\chi^2$ as the sum of squared differences between model and observed intensities, with uncertainties listed in Table 2 being weights,
\begin{equation}
\chi^2=\sum_{i}{\frac{[I_{\rm obs}(R_i)-I_0\widetilde{I}_{\rm mod}^{*}(R_i|r_0)
       -I_{\rm bkg}]^2}{\sigma_{i}^{2}}},
\end{equation}
in which a background $I_{\rm bkg}$ was also fitted.

Figure 1 displays the observed intensity profiles as a function of logarithmic projected radius and the best-fitting King model (solid red line) for each cluster. The observed data have been extinction corrected, following by a fitted $I_{\rm bkg}$ subtracted. The dashed blue lines represent the shape of the PSF for the WFC F606W or F814W filters. Most profiles of the sample clusters were well fitted by King model, except for those at the intermediate radii of GC3, GC7 and GC9. We checked the images, and found that the three clusters are very loose and there are several bright stars at the intermediate radii.

In Figure 1, open squares are {\sc ellipse} data points
included in the least-squares model fitting, and the crosses are points flagged as `DEP' or `BAD', which are not used to constrain the fit. In this paper, the {\sc ellipse} gives isophotal intensities for $15$ radii inside $R < 2$ pixels, however, all of them are derived from the same innermost $13$ pixels, meaning that the isophotal intensities are not statistically independent. So, to avoid excessive weighting of the central regions of clusters in the fits, we only used intensities at radii $R_{\rm min}$, $R_{\rm min}+(0.5,1.0,2.0)~{\rm pixels}$, or $R>2.5$ pixels as \citet{barmby07} used. In addition, we deleted some individual isophotes which deviated strongly from their neighbours or showed irregular features by hand.

\subsection{Basic Model Parameters}

Table 3 lists the basic parameters of 20 model fits to the sample clusters here. Column (1) gives the cluster name,
column (2) the detector/filter from which the observed data were derived. Column (3) gives the color
correction $(V-{\rm F606W})_0$ to transform native instrumental magnitudes to the standard $V$ scale.
The fourth column shows the number of points in the intensity profile that are flagged as `OK' in Table 2,
which were used for constraining the model fits. Column (5) is the fitting model which is always King model here. Column (6) gives the minimum $\chi^2$ obtained in the fits. Column (7) gives the best fitted background intensity. Column (8) gives the dimensionless central potential $W_0$ of the best-fitting model, defined as $W_0 \equiv -\phi(0)/\sigma_0^2$. Column (9) gives the concentration
$c \equiv \log(r_t/r_0)$. Column (10) gives the best-fit central surface brightness in the native bandpass
of the data. Column (11) and column (12) show the best model-fit scale radius $r_0$ in arcseconds and parsecs,
respectively, while the latter was obtained from the angular scale with the distance moduli given in Table 1.

Uncertainties for the fitted model parameters were estimated following $\triangle\chi^2 \le 1$ for
68\% confidence intervals. However, as \citet{barmby07} pointed out, because the formal error bars estimated by {\sc ellipse} for the isophotal intensities are
artificially small, the best-fit $\chi_{\rm min}^2$ can be exceedingly high ($\gg\!N_{\rm pts}$; the number of points used in the model fitting) even when a model fit is actually very good (see the values of $\chi_{\rm min}^2$ in Table 3), and  this would result in unrealistically small estimates of parameter uncertainties. So, we also re-scale the $\chi^2$ for all fitted models by a common factor chosen to make the global minimum $\chi_{\rm min}^2 = (N_{\rm pts}-4)$ as \citet{barmby07} did. Under this re-scaling, the global minimum reduced $\chi^2$ per degree of freedom is exactly one \citep[see][for details]{barmby07}.

\subsection{Derived Quantities}

Tables 4 and 5 give various derived parameters for the best-fitting models for each cluster
\citep[the details of their calculation are given by][]{maclau08}.

The contents of Table 4 are:\\
Column (4): $\log r_t$, the model tidal radius in parsecs.\\
Column (5): $\log R_c$, the projected core radius of the model fitting a cluster, which is defined as $I(R_c) = I_0/2$.\\
Column (6): $\log R_h$, the projected half-light, or effective, radius of a model, containing half the total
luminosity in projection.\\
Column (7): $\log (R_h/R_c)$, a measure of cluster concentration and relatively more model-independent than
$W_0$ or $c$.\\
Column (8): $\log I_0 = 0.4(26.402 - \mu_{V,0}$), the best-fit central ($R = 0$) luminosity surface density in the $V$ band, in units of $L_{\odot}$ pc$^{-2}$. The surface-brightness zero point of $26.402$ corresponds to a solar absolute magnitude $M_{V,\odot} = +4.83$\footnote{See http://www.ucolick.org/$\sim$cnaw/sun.html.}. $\mu_{V,0}$ is derived from applying the term $(V-{\rm F606W})_0$ to the fitted central surface brightness in column (10) of Table 3.\\
Column (9): $\log j_0$, the central ($r = 0$) luminosity volume density in the $V$ band in units of $L_{\odot}$ pc$^{-3}$.\\
Column (10): $\log L_V$, the $V$-band total integrated model luminosity, in units of $L_{\odot}$.\\
Column (11):~$V_{\rm tot} = 4.83 - 2.5 \log (L_V/L_{\odot}) + 5\log (D/10$ pc) is the total $V$-band magnitude of
a model cluster.\\
Column (12): $\log I_h \equiv \log (L_V/2{\pi}R_h^2$), the luminosity surface density averaged over the half-light/effective radius in the $V$ band, in units of $L_{\odot,V}$ pc$^{-2}$.

The uncertainties of these derived parameters were estimated (separately for each given model family)
by calculating them in each model that yields $\chi^2$ within 1 of the global minimum for a cluster, and then taking the differences between the extreme and best-fit values of the parameters \citep[see][for details]{mm05}.

The contents of Table 5 are:\\
Column (3): $\Upsilon_V^{\rm pop}$, the $V$-band mass-to-light ratio. The values of $\Upsilon_V^{\rm pop}$ were derived by applying the population synthesis models of \citet{bc03}, assuming a \citet{chab03} initial mass function (IMF) and age of 13 Gyr for all these clusters, with metallicities given in Table 1. Uncertainties of $\Upsilon_V^{\rm pop}$ include a $\pm2$ Gyr uncertainty in age, as well as a $\pm0.6$ uncertainty in [Fe/H].\\
Column (5): $\log M_{\rm tot} = \log \Upsilon_V^{\rm pop} + \log L_V$, the integrated cluster mass in solar units,
estimated from the total model luminosity $L_V$.\\
Column (6): $\log E_b$, the integrated binding energy in ergs, which is defined as $E_b \equiv -(1/2)
\int_0^{r_t} 4{\rm \pi}r^2\rho\phi{\rm d}r$.\\
Column (7): $\log \Sigma_0=\log \Upsilon_V^{\rm pop} + \log I_0$, the central surface mass density in units of $M_{\odot}$ pc$^{-2}$.\\
Column (8): $\log \rho_0 = \log \Upsilon_V^{\rm pop} + \log j_0$, the central volume density in units of $M_{\odot}$ pc$^{-3}$.\\
Column (9): $\log \Sigma_h = \log \Upsilon_V^{\rm pop} + \log I_h$, the surface mass density averaged over the
half-light/effective radius $R_h$, in units of $M_{\odot}$ pc$^{-2}$.\\
Column (10): $\log \sigma_{p,0}$, the predicted line-of-sight velocity dispersion at the cluster center in units of km s$^{-1}$.\\
Column (11): $\log \nu_{\rm esc,0}$, the predicted central ``escape'' velocity in units of km s$^{-1}$, with which a star can move out from the center of a cluster, which is defined as $\nu_{\rm esc,0}^2/\sigma_0^2 = 2[W_0 + GM_{\rm tot}/r_t\sigma_0^2]$.\\
Column (12): $\log t_{r,h}$, the two-body relaxation time at the model-projected half-mass radius in units of years,
estimated as $t_{r,h} = {\frac{2.06\times10^6 yr}{\ln(0.4M_{\rm tot}/m_{\star})}}{\frac{M_{\rm tot}^{1/2}R_h^{3/2}}{m_{\star}}}$.
Here, $m_{\star}$, the average stellar mass in a cluster is assumed to be 0.5$M_{\odot}$ \citep[see][for the details]{maclau08}\\
Column (13): $\log f_0 \equiv \log [\rho_0/(2\pi\sigma_c^2)^{3/2}$], the model's central phase-space density, in units of
$M_{\odot}$ pc$^{-3}$ (km s$^{-1}$)$^{-3}$.

The uncertainties of these derived dynamical quantities were estimated from their variations in each model
that yields $\chi^2$ within 1 of the global minimum for a cluster, as above, and combined in quadrature with the
uncertainties in $\Upsilon_V^{\rm pop}$.

\begin{figure*}
\figurenum{2}
\includegraphics[angle=0,scale=0.7]{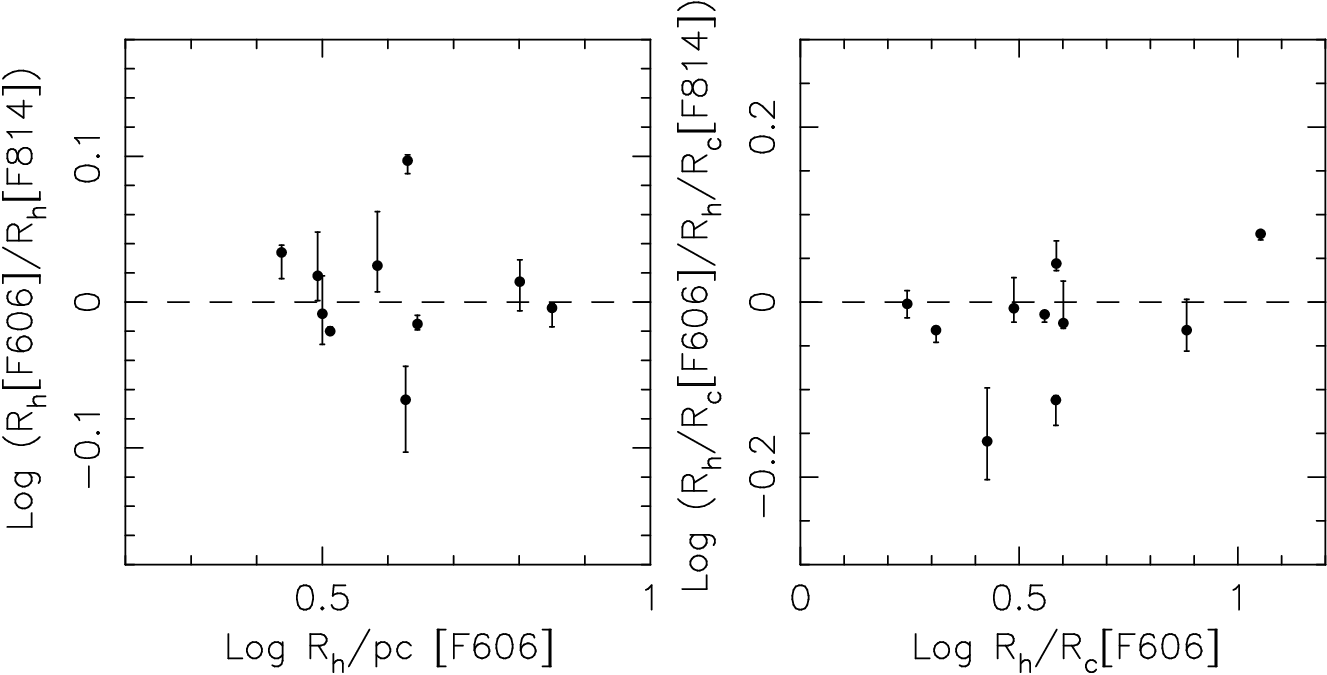}
\caption{Comparison of parameters for model fits to the sample clusters in both F606W and F814W filters. $left:$ projected half-light radius; $right:$ ratio of half-light to core radii.}
\label{fig2}
\end{figure*}

\subsection{Comparison of Results in the F606W and F814W Filters}

Model fits for the same cluster observed in different filters were compared to check whether there were systematic errors or color dependencies in the fits. Figure 2 shows the comparison of some parameters derived from fits to the sample clusters in both F606W and F814W filters. The left panel shows the comparison of projected half-light radius, while the right panel shows the ratio of half-light to core radii, all of which are from Table 4. The uncertainties for the parameters were also given in Figure 2. It is evident that the results between the two ACS bands are in good agreement. In following analysis, the F606W model fits were used for all the sample GCs.

\section{DISCUSSION}
\label{discussion.sec}

We combined the GC parameters derived here with those derived by King-model fits for clusters in the MW \citep{mm05} and M31 \citep{bhh02,barmby07,barmby09,huxor11} to form a large sample to look into the correlations between the parameters. The ellipticities and galactocentric distances for the MW GCs are from \citet{harris96} (2010 edition). For M31 GCs of \citet{bhh02} and \citet{barmby07} which were not observed in WFC F606W filter, the data of Space Telescope Imaging Spectrograph V-band or High Resolution Channel (HRC) F606W-band or HRC F555W-band or WFPC2 V-band are used, except for B082, of which the data of WFC F814W-band was used as \citet{barmby07} reported, since it was unsuccessfully fitted by King model in F606W filter. For clusters of \citet{barmby09} observed in WFPC2, the data of F439W- or F450W-band are used as \citet{barmby09} used, since these young clusters are dominated by blue stars and the measurements from the bluer filters are more preferred \citep[see][for the details]{barmby09}. \citet{huxor11} derived the structure parameters of 13 ECs based on $V$-band photometry (for INT) or $g$-band (for MegaCam) photometry. However, only metallicities of 4 ECs (HEC4, HEC5, HEC7 and HEC12) were determined by \citet{mackey06}. We derived integrated cluster mass for the four ECs, using the $V$-band absolute magnitude obtained by \citet{huxor11} and a mass-to-light ratio determined with the same approach here (see \S 3.2 for details).

\subsection{Ellipticity Distribution}

\begin{figure}
\figurenum{3}
\resizebox{\hsize}{!}{\rotatebox{0}{\includegraphics{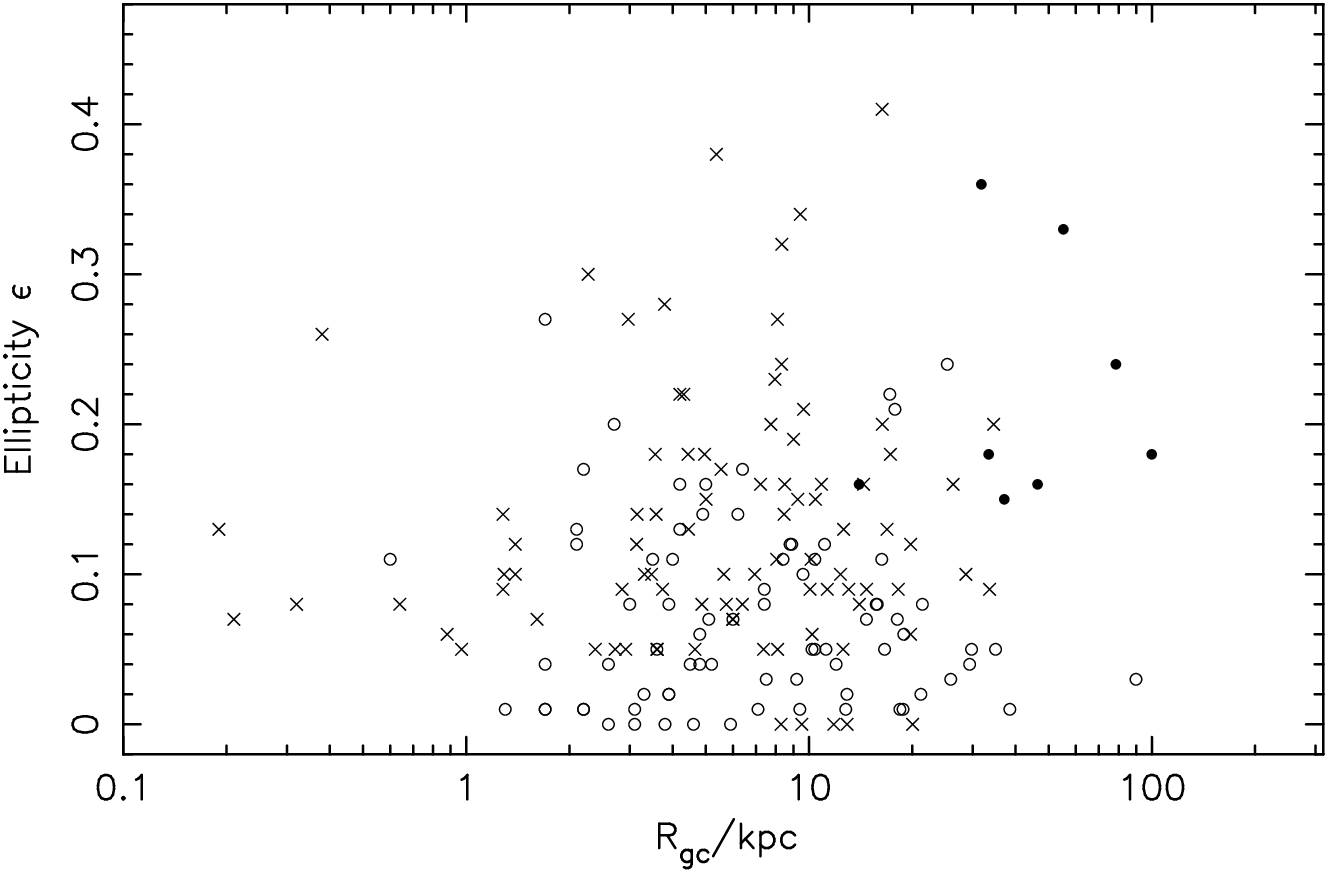}}}
\caption{Ellipticity vs. galactocentric distance for GCs in M31 from \citet{bhh02,barmby07} (crosses), in the outer halo of M31 from the present sample (filled circles) and the MW from \citet{harris96} (2010 edition) (open circles).}
\label{fig3}
\end{figure}

As noted by \citet{barmby07}, \citet{larsen01} listed several possible factors for the elongation of GCs: internal rotation, galaxy tides, cluster mergers, and ``remnant elongation'' from some clusters' former lives as dwarf galaxy nuclei. Cluster rotation is generally accepted to be a major factor for cluster flattening \citep{dp90}. However, \citet{vanden84} and \citet{vanden96} presented that the brightest GCs in both the MW and M31 are most flattened, which can be explained by the cluster mergers and ``remnant elongation''. In addition, as \citet{harris02} noted that dynamical models show that internal relaxation coupled to the external tides will in most situations drive a cluster toward a rounder shape over several relaxation times. \citet{harris02} presented that, the distributions of ellipticities for the M31 and NGC 5128 clusters and the old clusters in the Large Megallanic Cloud, are very similar, but different from the MW. With a large cluster sample, \citet{barmby07} also showed the distribution of ellipticities for clusters in the MW, M31 and NGC 5128, and found the distributions of ellipticities for M31 and NGC 5128 are not statistically different; both differ from the MW distribution in having few very round clusters. So, \citet{barmby07} concluded that there is no evidence that the overall galaxy environment is a major factor. \citet{bhh02} discussed about correlations of GC ellipticities with other properties in detail, and presented some explanations for these correlations combined with other authors' results \citep[see][and references therein]{bhh02}. In this paper, we show the distribution of ellipticity with galactocentric position for clusters in the MW and M31 in Figure 3, including the outer halo GCs in M31 from this study. A conclusion can be given that these outer halo GCs of M31 have larger ellipticities than most of GCs in M31 and the MW. These outer halo GCs lie at large projected radii than most sample clusters in \citet{bhh02,barmby07}, their large ellipticities may be due to galaxy tides coming from satellite dwarf galaxies of M31 or may be related to the apparently more vigorous accretion or merger history that M31 has experienced \citep[e.g.][]{Ibata05,Ibata07,McConnachie09,bekki10,hammer10,mackey10,huxor11}.

\begin{figure*}
\figurenum{4}
\resizebox{\hsize}{!}{\rotatebox{0}{\includegraphics{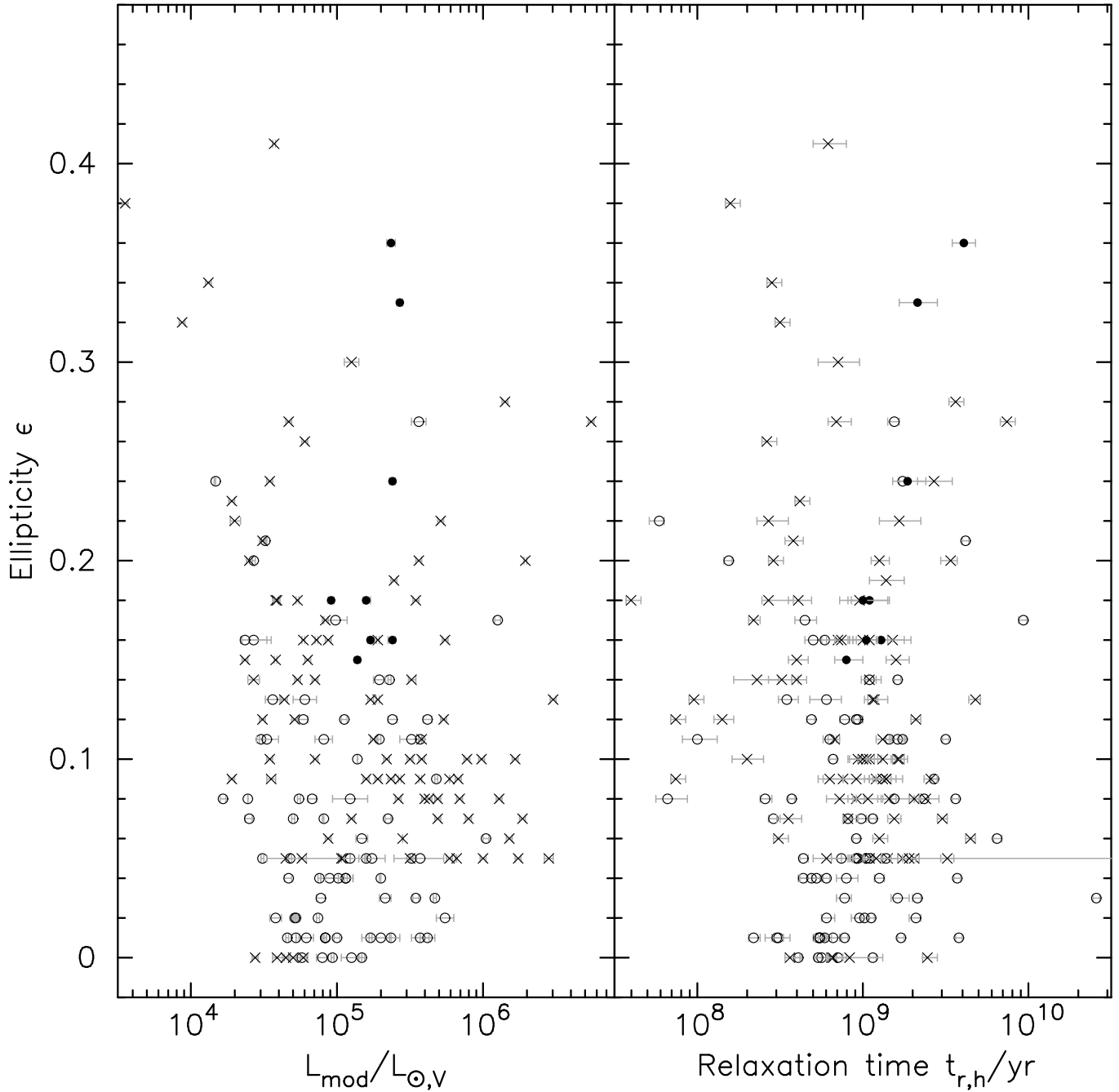}}}
\caption{Ellipticity as a function of luminosity ($left$) and half-mass relaxation time ($right$) for GCs. Symbols are as in Fig. 3.}
\label{fig4}
\end{figure*}

In order to show whether cluster ellipticities are caused by internal processes such as rotation or velocity anisotropy, \citet{barmby07} showed ellipticity as a function of luminosity and half-mass relaxation time for clusters in M31, NGC 5128 and the MW, since if it is true, relaxation through dynamical evolution should act to reduce any initial flattening \citep[see][and references therein]{barmby07}. These authors found a mild systematic decrease in ellipticity with increased luminosity, although considerable scatter, and no correlation of ellipticity with relaxation time is evident. So, \citet{barmby07} concluded that the observed distribution of GC ellipticity appears to be due to a number of factors. Figure 4 displays ellipticity as a function of model luminosity and half-mass relaxation time for clusters in the MW and M31, including the outer halo GCs in M31 studied here. It is evident that, when we add the data for the outer halo GCs in M31 obtained here, the conclusion of \citet{barmby07} will not evidently change, although the mild systematic decrease in ellipticity with increased luminosity nearly disappears.
In addition, we think that the larger ellipticities of the outer halo GCs of M31 than most of GCs in M31 and the MW may be due to galaxy tides coming from satellite dwarf galaxies of M31 or may be related to the apparently more vigorous accretion or merger history that M31 has experienced.

\subsection{Correlations with Position and Metallicity}

\begin{figure*}
\figurenum{5}
\resizebox{\hsize}{!}{\rotatebox{0}{\includegraphics{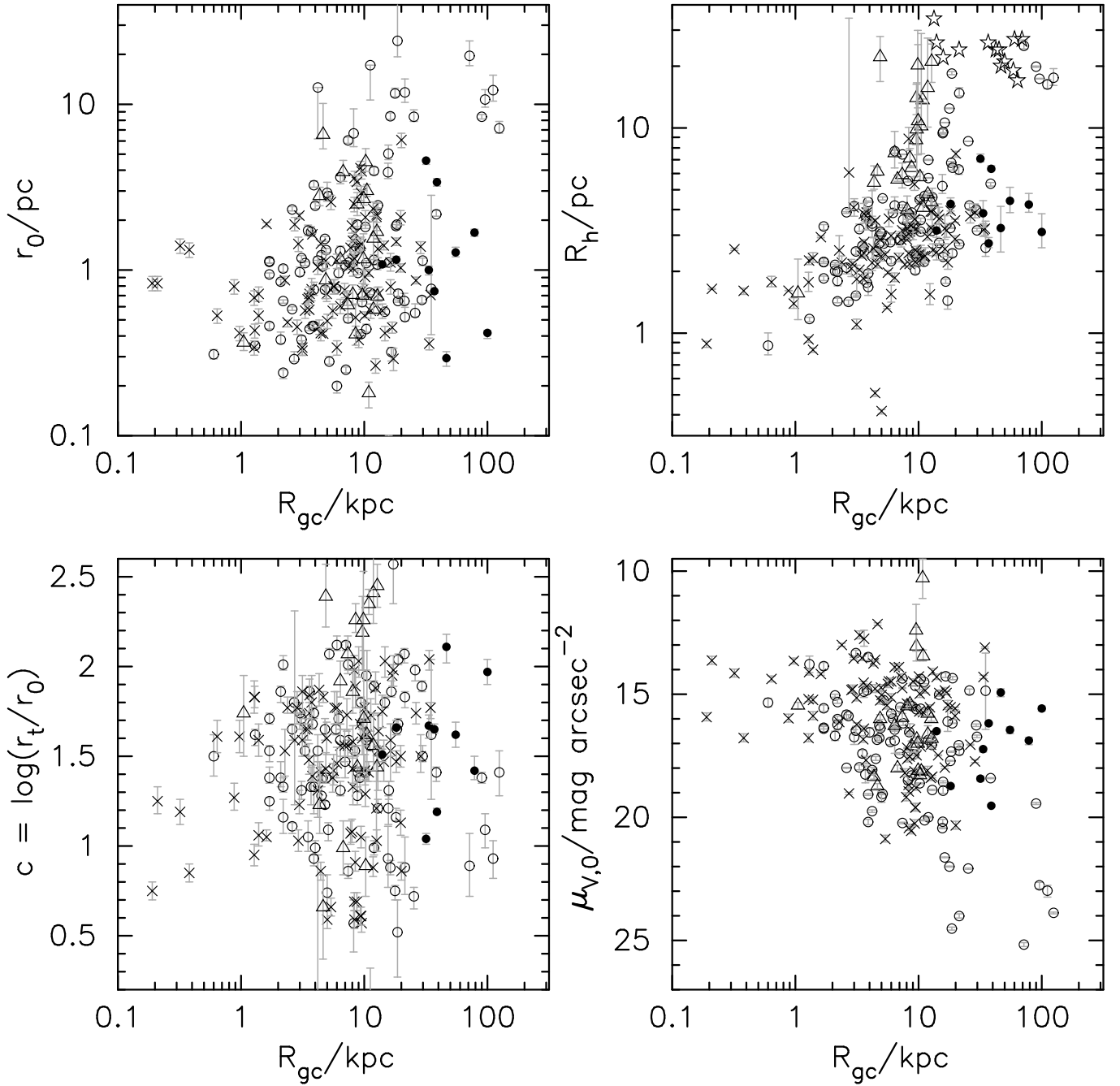}}}
\caption{Structural parameters vs. galactocentric distance $R_{\rm gc}$. The filled circles are GCs in the outer halo of M31 (this paper), the open circles are Galactic GCs \citep{mm05}, the crosses are M31 GCs \citep{bhh02,barmby07}, the open triangles are M31 young massive clusters \citep{barmby09}, the open stars are M31 ECs \citep{huxor11}.}
\label{fig5}
\end{figure*}

As \citet{barmby07} noted that, previous studies have shown that structures of the MW GCs are largely independent of galactocentric distances and metallicity, except for the correlation of half-light radius with galactocentric distances. \citet{bhh02} presented structural parameters as a function of galactocentric distance for clusters in M31 and the MW, and found that, there is no significant trend of $c$ with $R_{\rm gc}$, both $R_h$ and $r_0$ are correlated with $R_{\rm gc}$, and there is no clear correlation of $\mu_V(0)$ with $R_{\rm gc}$. The results of \citet{bhh02} are in agreement with ones obtained for GCs in the MW \citep[e.g.][]{maclau00} and NGC 5128 \citep[e.g.][]{harris02}. \citet{bhh02} concluded that the correlations of $R_h$ and $r_0$ with $R_{\rm gc}$ for GCs in both the MW and M31 are due to physical conditions at the time of cluster formation as suggested by \citet{vanden91} for MW GCs. \citet{barmby07} showed structural parameters as a function of galactocentric distance for GCs in the MW, the Magellanic Clouds and Fornax dwarf spheroidal, NGC 5128, and M31, and found similar results. In addition, \citet{mv05} showed that there is a clear trend of increasing $R_h$ with increasing $R_{\rm gc}$ for the Galactic GCs.
We should notice that the galactocentric distances are true three-dimensional distances for Galactic GCs and projected radii for M31 clusters.

Figure 5 shows structural parameters as a function of galactocentric distance $R_{\rm gc}$ for M31 outer halo GCs studied here, M31 young massive clusters \citep{barmby09}, MW globulars \citep{mm05}, M31 globulars \citep{bhh02,barmby07} and M31 ECs \citep{huxor11}. It is evident that, when we add the data for the outer halo GCs in M31 obtained here, the conclusion of \citet{bhh02} will not change, with the exception that the galactocentric distances of M31 clusters can reach to 100 kpc which are as distant as the MW clusters. In addition, it is true that M31 young massive clusters have larger $c$ and $R_h$ than old GCs at the same galactocentric distances.
For comparing with \citet{huxor11} (their Figure 9), we include the ECs of \citet{huxor11} in the upper-right panel of Figure 5, in which $R_h$ is versus $R_{\rm gc}$. It can be seen that, at large raddi (from 30 to 100 kpc) there are few GCs having $R_h$ in the range from 8 to 15 pc, which is in agreement with the finding of \citet{huxor11}.

\citet{bhh02} showed structural parameters as a function of [Fe/H] for clusters in M31 and the MW, and found that there is no correlation of metallicity with concentration $c$ or central surface brightness $\mu_V(0)$, but there does appear to be a correlation with size, as measured by $r_0$ or $R_h$, i.e. $r_0$ or $R_h$ deceases with increased metallicity. \citet{harris02} showed a different correlation of $r_h$ with [Fe/H] for GCs in NGC 5128, where $r_h$ is the half-mass radius, and reported that the correlation may be due to a selection effect because of a small sample.
\citet{barmby07} showed structural parameters as a function of [Fe/H] for GCs in the MW, the Magellanic Clouds and Fornax dwarf spheroidal, NGC 5128, and M31, and found that, no correlation of $c$ with [Fe/H] exists; a weak correlation of $R_h$ with [Fe/H] is present: $R_h$ deceases with increased metallicity, except for GCs in NGC 5128; there is a slight systematic increase of $\mu_{V,0}$ with [Fe/H].

Figure 6 plots structural parameters as a function of [Fe/H] for M31 outer halo GCs studied here, M31 young massive clusters \citep{barmby09}, MW globulars \citep{mm05}, and M31 globulars \citep{bhh02,barmby07} and M31 ECs \citep{huxor11}. It is evident that, the outer halo GCs of M31 fall in the same regions of parameter spaces of clusters in M31 and the MW. In addition, the conclusions of \citet{bhh02} and \citet{barmby07} do not change when adding the young massive clusters of \citet{barmby09} and the outer halo GCs here. We also include four ECs of \citet{huxor11} in the upper-right panel of Figure 6. An evident feature is that these four ECs are all metal-poor and have large $R_h$.

\begin{figure*}
\figurenum{6}
\resizebox{\hsize}{!}{\rotatebox{0}{\includegraphics{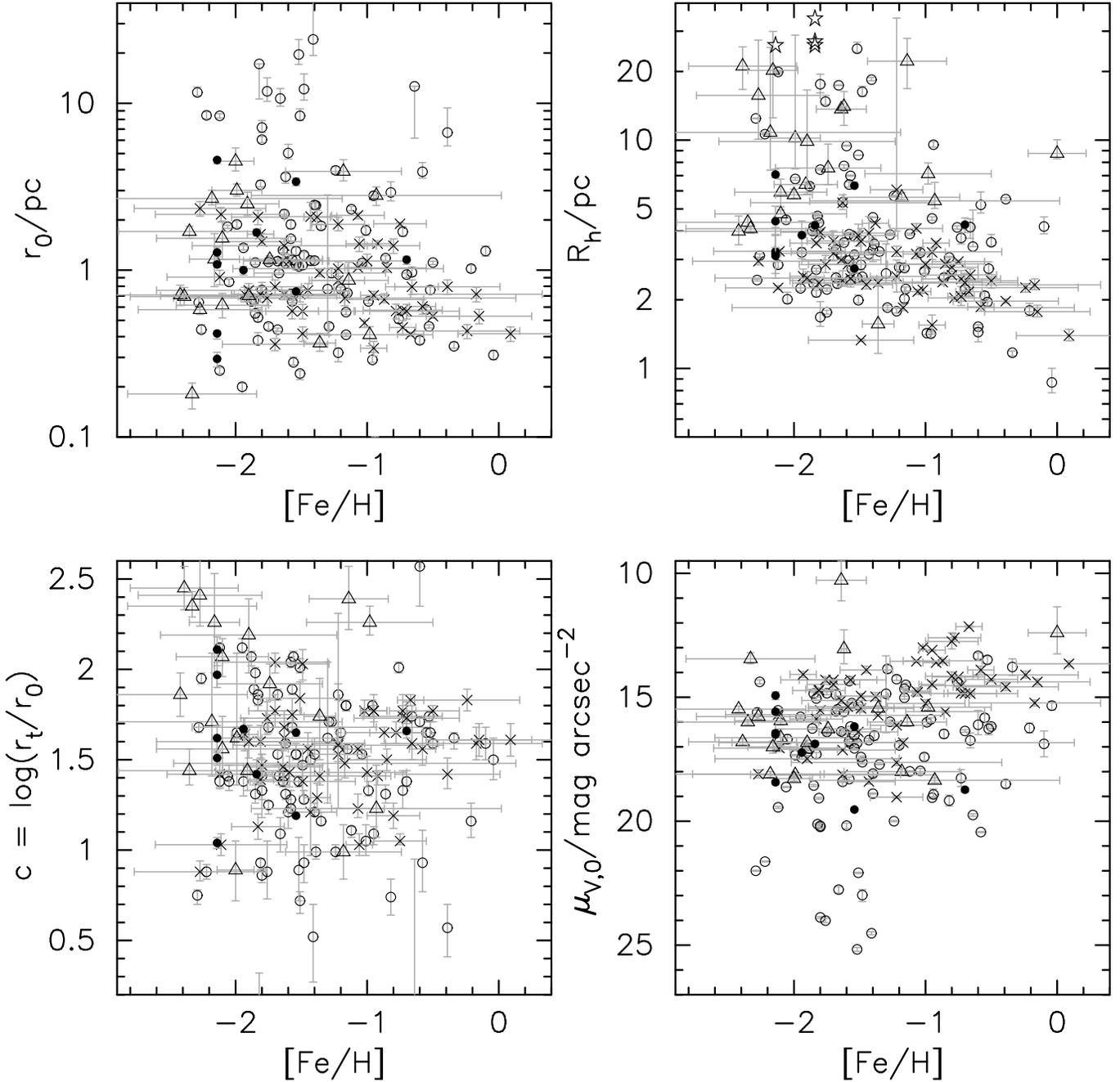}}}
\caption{Structural parameters as a function of [Fe/H]. Symbols are as in Fig. 5.}
\label{fig6}
\end{figure*}

Figure 7 plots structural parameters as a function of model mass $M_{\rm mod}$ for M31 outer halo GCs studied here, M31 young massive clusters \citep{barmby09}, MW globulars \citep{mm05}, M31 globulars \citep{bhh02,barmby07} and M31 ECs \citep{huxor11}. The properties of clusters in M31 and the MW fall in the same regions of parameter spaces, with the exception that, on average, the young massive clusters have larger sizes and higher concentrations than older clusters of the same mass \citep[see][for discussions in detail]{barmby09}. We also include four ECs of \citet{huxor11} in the upper-right panel of Figure 7. Three ECs have intermediate masses as well as the outer halo GCs of M31 studied here, while one EC (HEC12) has very low mass ($\sim2\times 10^4~{\rm M_\odot}$). \citet{barmby07} showed structural parameters as a function of model luminosity for GCs in M31, the MW, NGC 5128, the Magellanic Clouds, and the Fornax dSph, and found the properties of clusters in all six galaxies fall in the same regions of parameter spaces. The lower-right panel of Figure 7 shows one view of the fundamental plane, as defined by \citet{maclau00}. \citet{Djorgovski95} found a pair of bivariate correlations in MW GC parameters which imply the existence of a ``globular cluster fundamental plane'', similar to that expected if the cores were virialized structures. \citet{harris02} found that the NGC 5128 GCs describe a relation between binding energy and luminosity that even tighter than in the MW, which occupy the same extremely narrow region of the parametric ``fundamental plane'' as do their MW counterparts.

\begin{figure*}
\figurenum{7}
\resizebox{\hsize}{!}{\rotatebox{0}{\includegraphics{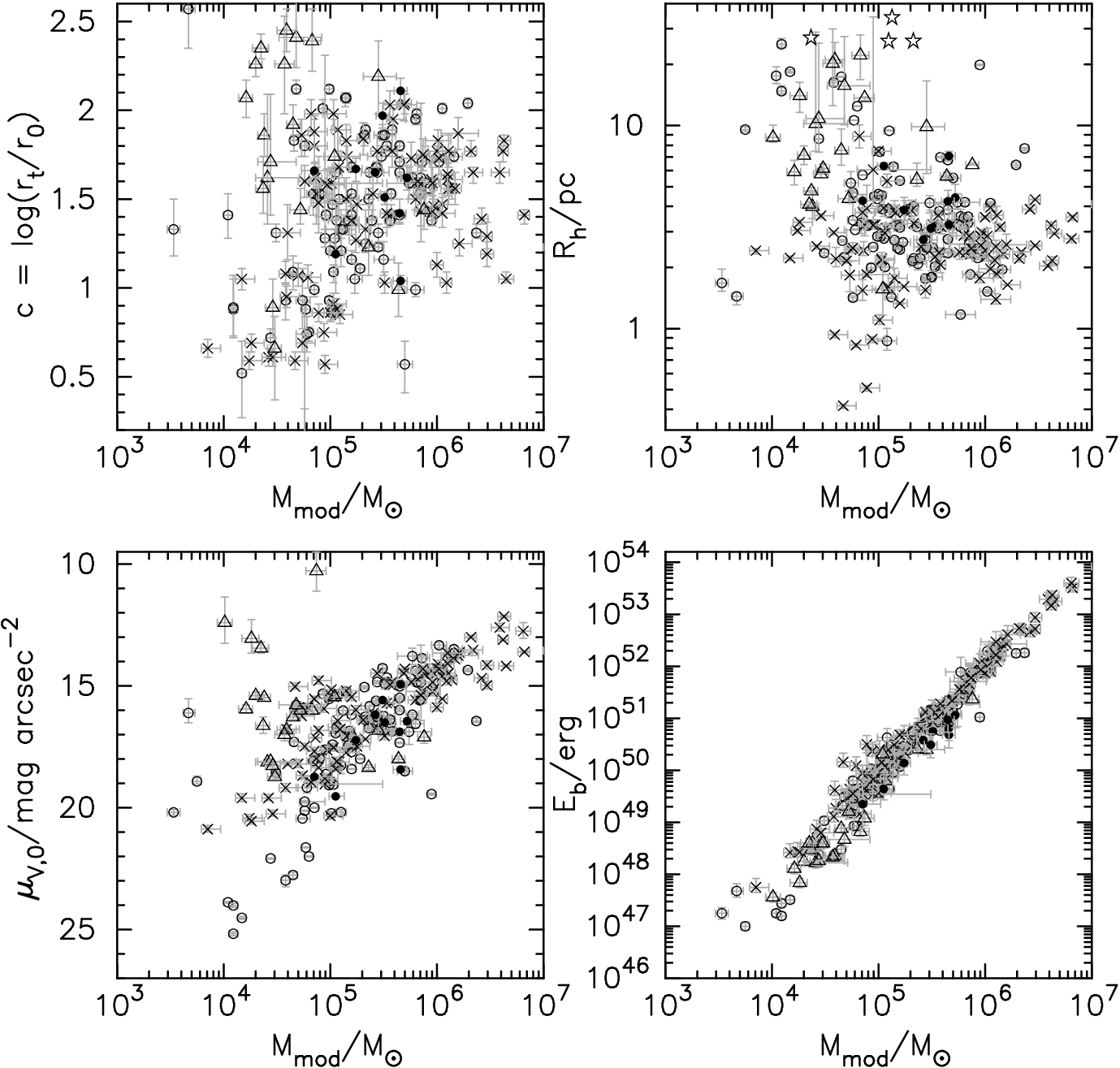}}}
\caption{Structural parameters as a function of model mass $M_{\rm mod}$. Symbols are as in Fig. 5.}
\label{fig7}
\end{figure*}

\section{SUMMARY}

GCs in the outer halo of M31 have recently been discovered in many surveys. We selected ten GCs (15 kpc $\lesssim R_p \lesssim$ 100 kpc) which have been studied by \citet{mackey07} based on the {\it HST} observations used in this paper. We measured surface brightness profiles for them using the {\it HST} images of \citet{mackey07}. Structural and dynamical parameters were derived by fitting the King model to the light profiles. We discussed the properties of the sample GCs here combined with GCs in the MW \citep{mm05} and clusters in M31 \citep{bhh02,barmby07,barmby09}. In general, the properties of the M31 and the Galactic clusters fall in the same regions of parameter spaces.

The outer halo GCs of M31, which lie at large projected radii, have larger ellipticities than most of GCs in M31 and the MW. Their large ellipticities may be due to galaxy tides coming from satellite dwarf galaxies of M31 or may be related to the apparently more vigorous accretion or
merger history that M31 has experienced. However, this conclusion remains to be checked because of the sample limitation. RBC V4.0 provides 39 GCs and 87 GC candidates which lie at $R_p > 20$ kpc. With more and more {\it HST} observations, structural and dynamical parameters for these clusters can be measured, which will provide a larger sample for discussion on the halo GCs.

The strong correlation of $E_b$ with model mass $M_{\rm mod}$ indicates a tight fundamental plane both for M31 and Galactic clusters, and no offset is apparent in the correlation between old and young clusters, especially including GCs in the outer halo of M31 studied here. This implies that some near-universal structural properties are present for clusters, regardless of their host environments, which is consistent with previous studies of \citet{barmby07,barmby09}.

\acknowledgments We would like to thank Dr. McLaughlin for his help in finishing this paper. He provide us a table including some parameters being model-dependent function of $W_0$ or $c$. An anonymous referee is thanked
for useful suggestions deriving from a careful and thorough reading of the original manuscript. This work was supported by the Chinese National Natural Science Foundation grant Nos. 10873016, and 10633020, and by the National Basic Research Program of China (973 Program) No. 2007CB815403.

\clearpage

\begin{table}
\centering
\small
\tabcolsep=2.5pt
\caption{Integrated Measurements for the 10 GCs in the M31 Halo.}
\label{t1.tab}
\begin{tabular}{ccccccccccc}
\tableline
\tableline
Name & $\epsilon^{a}$(F606W) & $\epsilon^{a}$(F814W) & $\theta^{b}$(F606W) & $\theta^{b}$(F814W) & $V$ & $I$ & ${R_{\rm gc}}$ & $(m-M)_0$ & $E(B-V)$ & ${\rm [Fe/H]}$ \\
     &     &     & (deg E of N)   & (deg E of N) & ({\sc vegamag}) & ({\sc vegamag}) & (kpc) &   &     &  \\
\hline
 GC1 & $0.16   \pm 0.09  $ & $0.23   \pm 0.14  $ & $-91    \pm 60    $ & $-86   \pm49    $ & 16.050   & 15.070   & 46.4     & 24.41    & 0.09     & $-2.14    $\\
 GC2 & $0.18   \pm 0.07  $ & $0.21   \pm 0.09  $ & $-134   \pm 36    $ & $-132  \pm38    $ & 16.980   & 16.040   & 33.4     & 24.32    & 0.08     & $-1.94    $\\
 GC3 & $0.36   \pm 0.25  $ & $...       $ & $-126   \pm 49    $ & $...      $ & 16.310   & 15.360   & 31.8     & 24.37    & 0.11     & $-2.14    $\\
 GC4 & $0.33   \pm 0.20  $ & $0.30   \pm 0.21  $ & $-139   \pm 61    $ & $-147  \pm57    $ & 15.760   & 14.680   & 55.2     & 24.35    & 0.09     & $-2.14    $\\
 GC5 & $0.24   \pm 0.16  $ & $0.15   \pm 0.08  $ & $-165   \pm 69    $ & $-165  \pm49    $ & 16.090   & 15.010   & 78.5     & 24.45    & 0.08     & $-1.84    $\\
 GC6{$^c$} & $0.16   \pm 0.12  $ & $0.32   \pm 0.26  $ & $-123   \pm 68    $ & $-116  \pm60    $ & 16.590   & 15.460   & 14.0     & 24.49    & 0.09     & $-2.14    $\\
 GC7 & $...       $ & $...       $ & $...       $ & $...      $ & 18.270   & 17.070   & 18.2     & 24.13    & 0.06     & $-0.70    $\\
 GC8 & $0.15   \pm 0.08  $ & $0.22   \pm 0.15  $ & $-183   \pm 43    $ & $-137  \pm55    $ & 16.720   & 15.680   & 37.1     & 24.43    & 0.09     & $-1.54    $\\
 GC9 & $...       $ & $0.35   \pm 0.16  $ & $...       $ & $-135  \pm43    $ & 17.780   & 16.710   & 38.9     & 24.22    & 0.15     & $-1.54    $\\
 GC10 & $0.18   \pm 0.13  $ & $0.13   \pm 0.05  $ & $-160   \pm 49    $ & $-152  \pm37    $ & 16.500   & 15.590   & 99.9     & 24.42    & 0.09     & $-2.14    $\\
 \tableline
\end{tabular}
\begin{minipage}[]{151mm}
\vspace{1mm}
{$^a$~$\epsilon=1-b/a$, while $a$ and $b$ are the lengths of the semimajor and semiminor axes, respectively.\\ $^b$~Position angle (P.A.)is measured in degrees anticlockwise from North.\\
$^c$~GC6 is called B298 in RBC V4.0.\\}
\end{minipage}
\end{table}

\begin{table}
\centering
\caption{The 20 F606W, F814W Intensity Profiles of the 10 GCs in the M31 Halo.}
\label{t2.tab}
\begin{tabular}{ccccccc}
\tableline
\tableline
Name & Detector & Filter & $R$ & $I$ & Uncertainty & Flag \\
  &  &  & (arcsec) & $L_{\odot}$~pc$^{-2}$ & $L_{\odot}$~pc$^{-2}$ & \\
(1) &  (2)  &  (3) &  (4) & (5) &  (6)  & (7) \\
\hline
 GC1    & WFC    & F606W    & 0.0260     & 31738.412    & 202.312      & OK    \\
 GC1    & WFC    & F606W    & 0.0287     & 31294.604    & 223.745      & DEP   \\
 GC1    & WFC    & F606W    & 0.0315     & 30807.689    & 243.173      & DEP   \\
 GC1    & WFC    & F606W    & 0.0347     & 30277.402    & 263.883      & DEP   \\
 GC1    & WFC    & F606W    & 0.0381     & 29710.863    & 294.703      & DEP   \\
 GC1    & WFC    & F606W    & 0.0420     & 29090.574    & 324.043      & DEP   \\
 GC1    & WFC    & F606W    & 0.0461     & 28412.213    & 351.340      & DEP   \\
 GC1    & WFC    & F606W    & 0.0508     & 27656.385    & 368.209      & DEP   \\
 GC1    & WFC    & F606W    & 0.0558     & 26734.781    & 357.023      & OK    \\
 GC1    & WFC    & F606W    & 0.0614     & 25634.490    & 362.977      & DEP   \\
 GC1    & WFC    & F606W    & 0.0676     & 24403.732    & 376.725      & DEP   \\
 \tableline
\end{tabular}
\tablecomments{Table 2 is published in its entirety
in the electronic edition of the Journal. Only a small portion is shown here, for guidance regarding its form and content. See text for description
of the FLAG column.}
 \end{table}

\begin{sidewaystable}
\centering
\small
\caption{Basic Parameters of the 20 Profiles of the 10 GCs in the M31 Halo.}
\label{t3.tab}
\begin{tabular}{cccccccccccc}
\tableline
\tableline
Name & Detector  & $(V-{\rm F606W})_0$ & $N_{\rm pts}$ & Model & $\chi_{\rm min}^2$ & $I_{\rm bkg}$ & $W_0$ & {$c$} &  $\mu_0$ & $\log r_0$ & $\log r_0$ \\
     &       &  (mag) &   &      &   & $L_{\odot}$~pc$^{-2}$ &    &    &     ${\rm (mag~arcsec^{-2})}$ & (arcsec) & (pc)\\
(1) &  (2)  &  (3) &  (4) & (5) &  (6)  & (7) &  (8)  &  (9) & (10) &  (11) &(12)\\
\hline
 GC1      & WFC/F606   & $0.189    \pm 0.050   $ & 57       & K66      & 3401.76    & $0.20     \pm 0.08    $ & $8.96    ^{+0.27    }_{-0.31   }$ & $2.11    ^{+0.07    }_{-0.09   }$ & $14.73   ^{+0.18    }_{-0.13   }$ & $-1.100  ^{+0.039   }_{-0.049  }$ & $-0.532  ^{+0.039   }_{-0.049  }$\\
     & WFC/F814   &           & 56       &     & 2083.76    & $0.00     \pm 0.13    $ & $8.72    ^{+0.27    }_{-0.32   }$ & $2.04    ^{+0.07    }_{-0.09   }$ & $14.35   ^{+0.20    }_{-0.17   }$ & $-1.000  ^{+0.032   }_{-0.048  }$ & $-0.432  ^{+0.032   }_{-0.048  }$\\
 GC2      & WFC/F606   & $0.183    \pm 0.050   $ & 53       & K66      & 686.35    & $0.20     \pm 0.12    $ & $7.49    ^{+0.20    }_{-0.22   }$ & $1.67    ^{+0.06    }_{-0.07   }$ & $17.05   ^{+0.11    }_{-0.06   }$ & $-0.550  ^{+0.022   }_{-0.030  }$ & $0.000   ^{+0.022   }_{-0.030  }$\\
     & WFC/F814   &           & 53       &     & 743.06    & $0.10     \pm 0.15    $ & $7.58    ^{+0.35    }_{-0.21   }$ & $1.70    ^{+0.11    }_{-0.06   }$ & $16.24   ^{+0.09    }_{-0.09   }$ & $-0.600  ^{+0.034   }_{-0.043  }$ & $-0.050  ^{+0.034   }_{-0.043  }$\\
 GC3      & WFC/F606   & $0.172    \pm 0.050   $ & 30       & K66      & 262.8     & $0.70     \pm 0.12    $ & $5.06    ^{+0.16    }_{-0.14   }$ & $1.04    ^{+0.03    }_{-0.03   }$ & $18.25   ^{+0.08    }_{-0.09   }$ & $0.100   ^{+0.007   }_{-0.023  }$ & $0.660   ^{+0.007   }_{-0.023  }$\\
     & WFC/F814   &           & 29       &     & 195.46     & $0.80     \pm 0.18    $ & $5.09    ^{+0.18    }_{-0.22   }$ & $1.05    ^{+0.04    }_{-0.05   }$ & $17.63   ^{+0.12    }_{-0.09   }$ & $0.100   ^{+0.012   }_{-0.012  }$ & $0.660   ^{+0.012   }_{-0.012  }$\\
 GC4      & WFC/F606   & $0.216    \pm 0.050   $ & 57       & K66      & 1378.26    & $0.60     \pm 0.16    $ & $7.31    ^{+0.21    }_{-0.24   }$ & $1.62    ^{+0.07    }_{-0.07   }$ & $16.24   ^{+0.14    }_{-0.12   }$ & $-0.450  ^{+0.031   }_{-0.027  }$ & $0.106   ^{+0.031   }_{-0.027  }$\\
     & WFC/F814   &           & 57       &     & 935.37    & $0.90     \pm 0.34    $ & $7.37    ^{+0.23    }_{-0.23   }$ & $1.64    ^{+0.07    }_{-0.07   }$ & $15.62   ^{+0.13    }_{-0.14   }$ & $-0.450  ^{+0.025   }_{-0.030  }$ & $0.106   ^{+0.025   }_{-0.030  }$\\
 GC5      & WFC/F606   & $0.222    \pm 0.050   $ & 49       & K66      & 687.47    & $1.00     \pm 0.31    $ & $6.62    ^{+0.28    }_{-0.20   }$ & $1.42    ^{+0.08    }_{-0.05   }$ & $16.65   ^{+0.16    }_{-0.11   }$ & $-0.350  ^{+0.020   }_{-0.022  }$ & $0.226   ^{+0.020   }_{-0.022  }$\\
     & WFC/F814   &           & 54       &     & 2066.58    & $0.40     \pm 1.03    $ & $7.43    ^{+0.27    }_{-0.33   }$ & $1.66    ^{+0.08    }_{-0.10   }$ & $15.87   ^{+0.16    }_{-0.09   }$ & $-0.450  ^{+0.029   }_{-0.060  }$ & $0.126   ^{+0.029   }_{-0.060  }$\\
 GC6      & WFC/F606   & $0.230    \pm 0.050   $ & 54       & K66      & 477.9     & $0.60     \pm 0.11    $ & $6.95    ^{+0.11    }_{-0.13   }$ & $1.51    ^{+0.03    }_{-0.04   }$ & $16.27   ^{+0.04    }_{-0.03   }$ & $-0.550  ^{+0.018   }_{-0.017  }$ & $0.034   ^{+0.018   }_{-0.017  }$\\
     & WFC/F814   &           & 54       &     & 404.46     & $0.90     \pm 0.25    $ & $6.98    ^{+0.22    }_{-0.26   }$ & $1.52    ^{+0.07    }_{-0.08   }$ & $15.67   ^{+0.09    }_{-0.12   }$ & $-0.550  ^{+0.015   }_{-0.026  }$ & $0.034   ^{+0.015   }_{-0.026  }$\\
 GC7      & WFC/F606   & $0.261    \pm 0.050   $ & 34       & K66      & 196.86     & $0.10     \pm 0.00    $ & $7.43    ^{+0.11    }_{-0.13   }$ & $1.66    ^{+0.04    }_{-0.04   }$ & $18.46   ^{+0.09    }_{-0.06   }$ & $-0.450  ^{+0.020   }_{-0.027  }$ & $0.062   ^{+0.020   }_{-0.027  }$\\
     & WFC/F814   &           & 34       &     & 229.16     & $0.20     \pm 0.00    $ & $7.22    ^{+0.14    }_{-0.11   }$ & $1.59    ^{+0.04    }_{-0.03   }$ & $17.18   ^{+0.08    }_{-0.05   }$ & $-0.500  ^{+0.021   }_{-0.049  }$ & $0.012   ^{+0.021   }_{-0.049  }$\\
 GC8      & WFC/F606   & $0.205    \pm 0.050   $ & 48       & K66      & 170.88     & $0.60     \pm 0.15    $ & $7.40    ^{+0.08    }_{-0.07   }$ & $1.65    ^{+0.03    }_{-0.02   }$ & $15.98   ^{+0.04    }_{-0.03   }$ & $-0.700  ^{+0.006   }_{-0.013  }$ & $-0.128  ^{+0.006   }_{-0.013  }$\\
     & WFC/F814   &           & 47       &     & 186.59     & $0.70     \pm 0.19    $ & $7.82    ^{+0.09    }_{-0.12   }$ & $1.78    ^{+0.03    }_{-0.04   }$ & $14.99   ^{+0.05    }_{-0.06   }$ & $-0.850  ^{+0.016   }_{-0.014  }$ & $-0.278  ^{+0.016   }_{-0.014  }$\\
 GC9      & WFC/F606   & $0.189    \pm 0.050   $ & 33       & K66      & 321.09     & $0.40     \pm 0.03    $ & $5.75    ^{+0.07    }_{-0.06   }$ & $1.19    ^{+0.02    }_{-0.01   }$ & $19.33   ^{+0.03    }_{-0.02   }$ & $0.000   ^{+0.022   }_{-0.024  }$ & $0.530   ^{+0.022   }_{-0.024  }$\\
     & WFC/F814   &           & 32       &     & 82.88     & $-0.30    \pm 0.09    $ & $6.02    ^{+0.20    }_{-0.16   }$ & $1.26    ^{+0.05    }_{-0.04   }$ & $18.63   ^{+0.13    }_{-0.07   }$ & $-0.050  ^{+0.017   }_{-0.015  }$ & $0.480   ^{+0.017   }_{-0.015  }$\\
 GC10     & WFC/F606   & $0.169    \pm 0.050   $ & 54       & K66      & 1433.16    & $1.20     \pm 0.14    $ & $8.45    ^{+0.25    }_{-0.24   }$ & $1.97    ^{+0.07    }_{-0.07   }$ & $15.42   ^{+0.10    }_{-0.09   }$ & $-0.950  ^{+0.015   }_{-0.033  }$ & $-0.380  ^{+0.015   }_{-0.033  }$\\
     & WFC/F814   &           & 54       &     & 1280.34    & $1.10     \pm 0.17    $ & $8.54    ^{+0.30    }_{-0.25   }$ & $1.99    ^{+0.08    }_{-0.07   }$ & $14.77   ^{+0.11    }_{-0.09   }$ & $-1.000  ^{+0.021   }_{-0.039  }$ & $-0.430  ^{+0.021   }_{-0.039  }$\\
\tableline
\end{tabular}
\end{sidewaystable}

\begin{sidewaystable}
\centering
\small
\caption{Derived Structural and Photometric Parameters from the 20 Profiles of the 10 GCs in the M31 Halo.}
\label{t4.tab}
\tabcolsep=2.5pt
\begin{tabular}{ccccccccccccc}
\tableline
\tableline
Name & Detector  & Model & $\log r_{\rm tid}$ & $\log R_c$ & $\log R_h$ & $\log R_h/R_c$ & $\log I_{\rm 0}$ & $\log j_{\rm 0}$ & $\log L_V$ & $V_{\rm tot}$  & $\log I_h$ & $<\mu_V>_h$\\
     &    &   & (pc) & (pc)  & (pc)  &   & $L_{\odot,V}$~pc$^{-2}$ &  $L_{\odot,V}$~pc$^{-3}$  & $L_{\odot,V}$  &  (mag)   &  $L_{\odot,V}$~pc$^{-2}$ &   ${\rm (mag~arcsec^{-2})}$  \\
(1) &  (2)  &  (3) &  (4) & (5) &  (6)  & (7) &  (8)  &  (9) & (10) &  (11) &(12) & (14) \\
\hline
 GC1      & WFC/F606   & K66      & $1.58    ^{+0.07    }_{-0.08   }$ & $-0.540  ^{+0.037   }_{-0.048  }$ & $0.512   ^{+0.106   }_{-0.117  }$ & $1.052   ^{+0.154   }_{-0.154  }$ & $4.59    ^{+0.06    }_{-0.07   }$ & $4.83    ^{+0.10    }_{-0.11   }$ & $5.38    ^{+0.02    }_{-0.03   }$ & $15.79   ^{+0.07    }_{-0.06   }$ & $3.56    ^{+0.21    }_{-0.19   }$ & $17.50   ^{+0.47    }_{-0.52   }$\\
     & WFC/F814   &     & $1.61    ^{+0.07    }_{-0.09   }$ & $-0.441  ^{+0.030   }_{-0.047  }$ & $0.532   ^{+0.109   }_{-0.117  }$ & $0.974   ^{+0.156   }_{-0.147  }$ &     & $4.73    ^{+0.10    }_{-0.10   }$ &     &     & $3.52    ^{+0.21    }_{-0.19   }$ & $17.61   ^{+0.48    }_{-0.52   }$\\
 GC2      & WFC/F606   & K66      & $1.67    ^{+0.06    }_{-0.07   }$ & $-0.017  ^{+0.020   }_{-0.028  }$ & $0.584   ^{+0.062   }_{-0.069  }$ & $0.601   ^{+0.091   }_{-0.089  }$ & $3.67    ^{+0.03    }_{-0.05   }$ & $3.38    ^{+0.06    }_{-0.07   }$ & $4.96    ^{+0.02    }_{-0.02   }$ & $16.74   ^{+0.05    }_{-0.05   }$ & $3.00    ^{+0.12    }_{-0.10   }$ & $18.91   ^{+0.26    }_{-0.29   }$\\
     & WFC/F814   &     & $1.65    ^{+0.11    }_{-0.06   }$ & $-0.066  ^{+0.032   }_{-0.041  }$ & $0.559   ^{+0.099   }_{-0.051  }$ & $0.625   ^{+0.139   }_{-0.083  }$ &     & $3.43    ^{+0.07    }_{-0.08   }$ &     &     & $3.05    ^{+0.08    }_{-0.18   }$ & $18.78   ^{+0.44    }_{-0.20   }$\\
 GC3      & WFC/F606   & K66      & $1.70    ^{+0.03    }_{-0.03   }$ & $0.607   ^{+0.004   }_{-0.019  }$ & $0.850   ^{+0.023   }_{-0.015  }$ & $0.244   ^{+0.043   }_{-0.019  }$ & $3.19    ^{+0.04    }_{-0.04   }$ & $2.28    ^{+0.06    }_{-0.04   }$ & $5.37    ^{+0.03    }_{-0.03   }$ & $15.77   ^{+0.08    }_{-0.08   }$ & $2.87    ^{+0.00    }_{-0.01   }$ & $19.21   ^{+0.03    }_{0.00    }$\\
     & WFC/F814   &     & $1.71    ^{+0.04    }_{-0.04   }$ & $0.608   ^{+0.007   }_{-0.008  }$ & $0.854   ^{+0.020   }_{-0.028  }$ & $0.246   ^{+0.028   }_{-0.035  }$ &     & $2.28    ^{+0.05    }_{-0.05   }$ &     &     & $2.87    ^{+0.02    }_{-0.01   }$ & $19.23   ^{+0.02    }_{-0.06   }$\\
 GC4      & WFC/F606   & K66      & $1.73    ^{+0.07    }_{-0.07   }$ & $0.087   ^{+0.028   }_{-0.025  }$ & $0.645   ^{+0.066   }_{-0.058  }$ & $0.558   ^{+0.091   }_{-0.086  }$ & $3.98    ^{+0.05    }_{-0.06   }$ & $3.59    ^{+0.08    }_{-0.09   }$ & $5.43    ^{+0.02    }_{-0.02   }$ & $15.60   ^{+0.06    }_{-0.05   }$ & $3.34    ^{+0.09    }_{-0.11   }$ & $18.04   ^{+0.28    }_{-0.23   }$\\
     & WFC/F814   &     & $1.74    ^{+0.07    }_{-0.07   }$ & $0.088   ^{+0.023   }_{-0.028  }$ & $0.660   ^{+0.060   }_{-0.054  }$ & $0.572   ^{+0.087   }_{-0.077  }$ &     & $3.59    ^{+0.08    }_{-0.08   }$ &     &     & $3.31    ^{+0.09    }_{-0.10   }$ & $18.12   ^{+0.25    }_{-0.21   }$\\
 GC5      & WFC/F606   & K66      & $1.64    ^{+0.08    }_{-0.05   }$ & $0.200   ^{+0.018   }_{-0.018  }$ & $0.627   ^{+0.054   }_{-0.038  }$ & $0.427   ^{+0.073   }_{-0.056  }$ & $3.81    ^{+0.05    }_{-0.07   }$ & $3.31    ^{+0.07    }_{-0.09   }$ & $5.38    ^{+0.02    }_{-0.02   }$ & $15.84   ^{+0.05    }_{-0.05   }$ & $3.32    ^{+0.06    }_{-0.09   }$ & $18.09   ^{+0.22    }_{-0.14   }$\\
     & WFC/F814   &     & $1.78    ^{+0.09    }_{-0.10   }$ & $0.108   ^{+0.025   }_{-0.057  }$ & $0.694   ^{+0.077   }_{-0.074  }$ & $0.586   ^{+0.134   }_{-0.100  }$ &     & $3.40    ^{+0.11    }_{-0.09   }$ &     &     & $3.19    ^{+0.13    }_{-0.13   }$ & $18.43   ^{+0.33    }_{-0.32   }$\\
 GC6      & WFC/F606   & K66      & $1.55    ^{+0.03    }_{-0.04   }$ & $0.012   ^{+0.017   }_{-0.016  }$ & $0.500   ^{+0.030   }_{-0.033  }$ & $0.488   ^{+0.045   }_{-0.050  }$ & $3.96    ^{+0.02    }_{-0.02   }$ & $3.64    ^{+0.04    }_{-0.04   }$ & $5.23    ^{+0.02    }_{-0.02   }$ & $16.26   ^{+0.05    }_{-0.05   }$ & $3.43    ^{+0.05    }_{-0.04   }$ & $17.83   ^{+0.10    }_{-0.12   }$\\
     & WFC/F814   &     & $1.55    ^{+0.07    }_{-0.07   }$ & $0.012   ^{+0.012   }_{-0.024  }$ & $0.508   ^{+0.056   }_{-0.054  }$ & $0.495   ^{+0.080   }_{-0.066  }$ &     & $3.64    ^{+0.05    }_{-0.04   }$ &     &     & $3.41    ^{+0.09    }_{-0.09   }$ & $17.87   ^{+0.23    }_{-0.22   }$\\
 GC7      & WFC/F606   & K66      & $1.72    ^{+0.04    }_{-0.04   }$ & $0.044   ^{+0.019   }_{-0.026  }$ & $0.630   ^{+0.029   }_{-0.035  }$ & $0.585   ^{+0.056   }_{-0.054  }$ & $3.07    ^{+0.03    }_{-0.04   }$ & $2.72    ^{+0.06    }_{-0.06   }$ & $4.46    ^{+0.03    }_{-0.03   }$ & $17.80   ^{+0.06    }_{-0.06   }$ & $2.40    ^{+0.04    }_{-0.03   }$ & $20.39   ^{+0.08    }_{-0.11   }$\\
     & WFC/F814   &     & $1.60    ^{+0.04    }_{-0.03   }$ & $-0.008  ^{+0.020   }_{-0.048  }$ & $0.533   ^{+0.033   }_{-0.026  }$ & $0.541   ^{+0.082   }_{-0.046  }$ &     & $2.78    ^{+0.08    }_{-0.06   }$ &     &     & $2.60    ^{+0.03    }_{-0.04   }$ & $19.91   ^{+0.10    }_{-0.07   }$\\
 GC8      & WFC/F606   & K66      & $1.52    ^{+0.03    }_{-0.02   }$ & $-0.146  ^{+0.005   }_{-0.013  }$ & $0.438   ^{+0.028   }_{-0.021  }$ & $0.584   ^{+0.041   }_{-0.026  }$ & $4.09    ^{+0.02    }_{-0.03   }$ & $3.93    ^{+0.04    }_{-0.03   }$ & $5.14    ^{+0.02    }_{-0.02   }$ & $16.40   ^{+0.05    }_{-0.05   }$ & $3.47    ^{+0.02    }_{-0.04   }$ & $17.73   ^{+0.09    }_{-0.05   }$\\
     & WFC/F814   &     & $1.50    ^{+0.03    }_{-0.04   }$ & $-0.293  ^{+0.015   }_{-0.013  }$ & $0.404   ^{+0.033   }_{-0.039  }$ & $0.696   ^{+0.046   }_{-0.055  }$ &     & $4.07    ^{+0.04    }_{-0.04   }$ &     &     & $3.54    ^{+0.06    }_{-0.05   }$ & $17.55   ^{+0.11    }_{-0.15   }$\\
 GC9      & WFC/F606   & K66      & $1.72    ^{+0.02    }_{-0.01   }$ & $0.491   ^{+0.021   }_{-0.023  }$ & $0.801   ^{+0.015   }_{-0.008  }$ & $0.310   ^{+0.038   }_{-0.028  }$ & $2.75    ^{+0.02    }_{-0.02   }$ & $1.96    ^{+0.05    }_{-0.04   }$ & $4.77    ^{+0.02    }_{-0.02   }$ & $17.12   ^{+0.05    }_{-0.06   }$ & $2.37    ^{+-0.01   }_{-0.01   }$ & $20.47   ^{+0.02    }_{0.01    }$\\
     & WFC/F814   &     & $1.74    ^{+0.05    }_{-0.04   }$ & $0.446   ^{+0.015   }_{-0.012  }$ & $0.787   ^{+0.030   }_{-0.028  }$ & $0.342   ^{+0.042   }_{-0.042  }$ &     & $2.00    ^{+0.03    }_{-0.04   }$ &     &     & $2.40    ^{+0.03    }_{-0.04   }$ & $20.40   ^{+0.09    }_{-0.09   }$\\
 GC10     & WFC/F606   & K66      & $1.59    ^{+0.07    }_{-0.07   }$ & $-0.391  ^{+0.014   }_{-0.032  }$ & $0.493   ^{+0.089   }_{-0.077  }$ & $0.883   ^{+0.121   }_{-0.090  }$ & $4.33    ^{+0.04    }_{-0.04   }$ & $4.41    ^{+0.07    }_{-0.06   }$ & $5.20    ^{+0.03    }_{-0.02   }$ & $16.24   ^{+0.06    }_{-0.07   }$ & $3.42    ^{+0.13    }_{-0.15   }$ & $17.85   ^{+0.38    }_{-0.33   }$\\
     & WFC/F814   &     & $1.56    ^{+0.09    }_{-0.07   }$ & $-0.440  ^{+0.020   }_{-0.037  }$ & $0.475   ^{+0.119   }_{-0.094  }$ & $0.915   ^{+0.156   }_{-0.114  }$ &     & $4.46    ^{+0.08    }_{-0.06   }$ &     &     & $3.46    ^{+0.17    }_{-0.21   }$ & $17.76   ^{+0.53    }_{-0.41   }$\\
\tableline
\end{tabular}
\end{sidewaystable}

\begin{sidewaystable}
\centering
\small
\caption{Derived Dynamical Parameters from the 20 Profiles of the 10 GCs in the M31 Halo.}
\label{t5.tab}
\tabcolsep=2.5pt
\begin{tabular}{ccccccccccccc}
\tableline
\tableline
Name & Detector & $\Upsilon_V^{\rm pop}$  & Model & $\log M_{\rm tot}$ & $\log E_b$ & $\log \Sigma_{\rm 0}$ & $\log \rho_{\rm 0}$ & $\log \Sigma_h$ & $\log \sigma_{p,0}$ & $\log \nu_{\rm esc,0}$ & $\log t_{r,h}$ & $\log f_{\rm 0}$\\
     &    & $M_{\odot}~L_{\odot,V}^{-1}$  & & $M_{\odot}$ & (erg)  & $M_{\odot}$~pc$^{-2}$ & $M_{\odot}$~pc$^{-3}$   &  $M_{\odot}$~pc$^{-2}$ &  (km s$^{-1}$)  & (km s$^{-1}$)  & yr & $M_{\odot}$~(pc~km~s$^{-1})^{-3})$ \\
(1) &  (2)  &  (3) &  (4) & (5) &  (6)  & (7) &  (8)  &  (9) & (10) &  (11) &(12) & (13) \\
\hline
 GC1      & WFC/F606   & $1.918   ^{+0.244   }_{-0.237  }$ & K66      & $5.66    ^{+0.06    }_{-0.06   }$ & $50.68   ^{+0.21    }_{-0.25   }$ & $4.88    ^{+0.08    }_{-0.09   }$ & $5.11    ^{+0.12    }_{-0.13   }$ & $3.84    ^{+0.20    }_{-0.22   }$ & $0.906   ^{+0.026   }_{-0.034  }$ & $1.539   ^{+0.028   }_{-0.038  }$ & $9.11    ^{+0.18    }_{-0.20   }$ & $1.183   ^{+0.098   }_{-0.065  }$\\
     & WFC/F814   & $1.918   ^{+0.244   }_{-0.237  }$ &     &     & $50.93   ^{+0.21    }_{-0.26   }$ & $4.88    ^{+0.08    }_{-0.09   }$ & $5.01    ^{+0.12    }_{-0.12   }$ & $3.80    ^{+0.20    }_{-0.21   }$ & $0.956   ^{+0.026   }_{-0.036  }$ & $1.584   ^{+0.028   }_{-0.041  }$ & $9.14    ^{+0.18    }_{-0.19   }$ & $0.933   ^{+0.096   }_{-0.050  }$\\
 GC2      & WFC/F606   & $1.889   ^{+0.257   }_{-0.226  }$ & K66      & $5.24    ^{+0.06    }_{-0.06   }$ & $50.14   ^{+0.22    }_{-0.23   }$ & $3.94    ^{+0.06    }_{-0.07   }$ & $3.66    ^{+0.08    }_{-0.09   }$ & $3.27    ^{+0.12    }_{-0.13   }$ & $0.704   ^{+0.028   }_{-0.031  }$ & $1.305   ^{+0.028   }_{-0.034  }$ & $9.04    ^{+0.12    }_{-0.13   }$ & $0.325   ^{+0.064   }_{-0.040  }$\\
     & WFC/F814   & $1.889   ^{+0.257   }_{-0.226  }$ &     &     & $50.00   ^{+0.22    }_{-0.22   }$ & $3.94    ^{+0.06    }_{-0.07   }$ & $3.71    ^{+0.09    }_{-0.10   }$ & $3.32    ^{+0.18    }_{-0.10   }$ & $0.679   ^{+0.028   }_{-0.029  }$ & $1.282   ^{+0.028   }_{-0.031  }$ & $9.00    ^{+0.17    }_{-0.10   }$ & $0.449   ^{+0.094   }_{-0.064  }$\\
 GC3      & WFC/F606   & $1.918   ^{+0.244   }_{-0.237  }$ & K66      & $5.66    ^{+0.06    }_{-0.07   }$ & $50.81   ^{+0.21    }_{-0.24   }$ & $3.47    ^{+0.07    }_{-0.07   }$ & $2.56    ^{+0.08    }_{-0.07   }$ & $3.16    ^{+0.05    }_{-0.06   }$ & $0.784   ^{+0.028   }_{-0.034  }$ & $1.331   ^{+0.030   }_{-0.035  }$ & $9.61    ^{+0.07    }_{-0.07   }$ & $-1.041  ^{+0.040   }_{-0.030  }$\\
     & WFC/F814   & $1.918   ^{+0.244   }_{-0.237  }$ &     &     & $50.82   ^{+0.22    }_{-0.24   }$ & $3.47    ^{+0.07    }_{-0.07   }$ & $2.56    ^{+0.07    }_{-0.07   }$ & $3.15    ^{+0.05    }_{-0.06   }$ & $0.784   ^{+0.031   }_{-0.033  }$ & $1.332   ^{+0.033   }_{-0.035  }$ & $9.62    ^{+0.07    }_{-0.08   }$ & $-1.042  ^{+0.026   }_{-0.029  }$\\
 GC4      & WFC/F606   & $1.918   ^{+0.244   }_{-0.237  }$ & K66      & $5.72    ^{+0.06    }_{-0.06   }$ & $51.06   ^{+0.22    }_{-0.24   }$ & $4.26    ^{+0.07    }_{-0.08   }$ & $3.87    ^{+0.09    }_{-0.10   }$ & $3.63    ^{+0.12    }_{-0.11   }$ & $0.914   ^{+0.029   }_{-0.032  }$ & $1.511   ^{+0.031   }_{-0.035  }$ & $9.33    ^{+0.12    }_{-0.11   }$ & $-0.096  ^{+0.048   }_{-0.054  }$\\
     & WFC/F814   & $1.918   ^{+0.244   }_{-0.237  }$ &     &     & $51.07   ^{+0.21    }_{-0.24   }$ & $4.26    ^{+0.07    }_{-0.08   }$ & $3.87    ^{+0.09    }_{-0.10   }$ & $3.60    ^{+0.11    }_{-0.10   }$ & $0.914   ^{+0.028   }_{-0.033  }$ & $1.513   ^{+0.031   }_{-0.036  }$ & $9.35    ^{+0.11    }_{-0.11   }$ & $-0.097  ^{+0.054   }_{-0.043  }$\\
 GC5      & WFC/F606   & $1.881   ^{+0.275   }_{-0.223  }$ & K66      & $5.65    ^{+0.06    }_{-0.06   }$ & $50.97   ^{+0.25    }_{-0.24   }$ & $4.09    ^{+0.08    }_{-0.09   }$ & $3.58    ^{+0.09    }_{-0.10   }$ & $3.60    ^{+0.11    }_{-0.08   }$ & $0.885   ^{+0.033   }_{-0.037  }$ & $1.467   ^{+0.036   }_{-0.040  }$ & $9.27    ^{+0.11    }_{-0.09   }$ & $-0.302  ^{+0.039   }_{-0.030  }$\\
     & WFC/F814   & $1.881   ^{+0.275   }_{-0.223  }$ &     &     & $50.79   ^{+0.24    }_{-0.24   }$ & $4.09    ^{+0.08    }_{-0.09   }$ & $3.68    ^{+0.12    }_{-0.11   }$ & $3.46    ^{+0.15    }_{-0.14   }$ & $0.838   ^{+0.030   }_{-0.035  }$ & $1.438   ^{+0.030   }_{-0.040  }$ & $9.38    ^{+0.14    }_{-0.13   }$ & $-0.061  ^{+0.126   }_{-0.044  }$\\
 GC6      & WFC/F606   & $1.918   ^{+0.244   }_{-0.237  }$ & K66      & $5.51    ^{+0.06    }_{-0.06   }$ & $50.75   ^{+0.21    }_{-0.23   }$ & $4.24    ^{+0.06    }_{-0.06   }$ & $3.93    ^{+0.07    }_{-0.07   }$ & $3.71    ^{+0.07    }_{-0.07   }$ & $0.868   ^{+0.026   }_{-0.029  }$ & $1.458   ^{+0.027   }_{-0.029  }$ & $9.02    ^{+0.07    }_{-0.08   }$ & $0.096   ^{+0.039   }_{-0.042  }$\\
     & WFC/F814   & $1.918   ^{+0.244   }_{-0.237  }$ &     &     & $50.76   ^{+0.21    }_{-0.23   }$ & $4.24    ^{+0.06    }_{-0.06   }$ & $3.93    ^{+0.07    }_{-0.07   }$ & $3.70    ^{+0.11    }_{-0.11   }$ & $0.868   ^{+0.026   }_{-0.029  }$ & $1.458   ^{+0.026   }_{-0.031  }$ & $9.03    ^{+0.11    }_{-0.11   }$ & $0.096   ^{+0.057   }_{-0.036  }$\\
 GC7      & WFC/F606   & $2.441   ^{+1.121   }_{-0.573  }$ & K66      & $4.85    ^{+0.17    }_{-0.12   }$ & $49.35   ^{+0.66    }_{-0.47   }$ & $3.46    ^{+0.17    }_{-0.12   }$ & $3.11    ^{+0.17    }_{-0.13   }$ & $2.79    ^{+0.17    }_{-0.12   }$ & $0.492   ^{+0.082   }_{-0.059  }$ & $1.093   ^{+0.082   }_{-0.060  }$ & $8.94    ^{+0.17    }_{-0.13   }$ & $0.413   ^{+0.097   }_{-0.065  }$\\
     & WFC/F814   & $2.441   ^{+1.121   }_{-0.573  }$ &     &     & $49.16   ^{+0.66    }_{-0.47   }$ & $3.46    ^{+0.17    }_{-0.12   }$ & $3.16    ^{+0.18    }_{-0.13   }$ & $2.98    ^{+0.17    }_{-0.12   }$ & $0.467   ^{+0.083   }_{-0.059  }$ & $1.062   ^{+0.082   }_{-0.060  }$ & $8.80    ^{+0.17    }_{-0.13   }$ & $0.540   ^{+0.135   }_{-0.065  }$\\
 GC8      & WFC/F606   & $1.897   ^{+0.387   }_{-0.215  }$ & K66      & $5.42    ^{+0.08    }_{-0.06   }$ & $50.58   ^{+0.32    }_{-0.21   }$ & $4.36    ^{+0.08    }_{-0.06   }$ & $4.21    ^{+0.09    }_{-0.06   }$ & $3.75    ^{+0.09    }_{-0.06   }$ & $0.849   ^{+0.041   }_{-0.028  }$ & $1.449   ^{+0.041   }_{-0.029  }$ & $8.90    ^{+0.10    }_{-0.07   }$ & $0.436   ^{+0.046   }_{-0.026  }$\\
     & WFC/F814   & $1.897   ^{+0.387   }_{-0.215  }$ &     &     & $50.20   ^{+0.32    }_{-0.21   }$ & $4.36    ^{+0.08    }_{-0.06   }$ & $4.35    ^{+0.09    }_{-0.07   }$ & $3.82    ^{+0.09    }_{-0.08   }$ & $0.775   ^{+0.041   }_{-0.027  }$ & $1.384   ^{+0.041   }_{-0.027  }$ & $8.84    ^{+0.10    }_{-0.09   }$ & $0.808   ^{+0.046   }_{-0.037  }$\\
 GC9      & WFC/F606   & $1.897   ^{+0.387   }_{-0.215  }$ & K66      & $5.05    ^{+0.08    }_{-0.06   }$ & $49.64   ^{+0.32    }_{-0.21   }$ & $3.03    ^{+0.08    }_{-0.06   }$ & $2.23    ^{+0.09    }_{-0.07   }$ & $2.65    ^{+0.08    }_{-0.05   }$ & $0.504   ^{+0.040   }_{-0.026  }$ & $1.066   ^{+0.040   }_{-0.026  }$ & $9.28    ^{+0.09    }_{-0.06   }$ & $-0.516  ^{+0.063   }_{-0.049  }$\\
     & WFC/F814   & $1.897   ^{+0.387   }_{-0.215  }$ &     &     & $49.53   ^{+0.32    }_{-0.21   }$ & $3.03    ^{+0.08    }_{-0.06   }$ & $2.28    ^{+0.09    }_{-0.06   }$ & $2.68    ^{+0.09    }_{-0.06   }$ & $0.480   ^{+0.041   }_{-0.026  }$ & $1.049   ^{+0.041   }_{-0.027  }$ & $9.26    ^{+0.10    }_{-0.07   }$ & $-0.398  ^{+0.046   }_{-0.039  }$\\
 GC10     & WFC/F606   & $1.918   ^{+0.244   }_{-0.237  }$ & K66      & $5.49    ^{+0.06    }_{-0.06   }$ & $50.49   ^{+0.21    }_{-0.25   }$ & $4.61    ^{+0.07    }_{-0.07   }$ & $4.69    ^{+0.09    }_{-0.08   }$ & $3.70    ^{+0.16    }_{-0.14   }$ & $0.848   ^{+0.026   }_{-0.032  }$ & $1.470   ^{+0.028   }_{-0.035  }$ & $9.00    ^{+0.15    }_{-0.14   }$ & $0.938   ^{+0.067   }_{-0.032  }$\\
     & WFC/F814   & $1.918   ^{+0.244   }_{-0.237  }$ &     &     & $50.36   ^{+0.21    }_{-0.24   }$ & $4.61    ^{+0.07    }_{-0.07   }$ & $4.74    ^{+0.09    }_{-0.09   }$ & $3.74    ^{+0.22    }_{-0.18   }$ & $0.823   ^{+0.026   }_{-0.031  }$ & $1.447   ^{+0.027   }_{-0.033  }$ & $8.98    ^{+0.20    }_{-0.16   }$ & $1.063   ^{+0.080   }_{-0.042  }$\\
\tableline
\end{tabular}
\end{sidewaystable}

 \end{document}